\documentclass[manuscript]{acmart}

\AtBeginDocument{%
}

\setcopyright{none}
\renewcommand\footnotetextcopyrightpermission[1]{}

\usepackage{booktabs}
\usepackage{tabularx}
\usepackage{array}
\usepackage{xcolor}
\usepackage{graphicx}
\usepackage{placeins}
\usepackage{multirow}
\usepackage{pifont}
\usepackage{makecell}
\usepackage{enumitem}
\usepackage{microtype}
\usepackage{hyperref}
\usepackage{CJKutf8}

\usepackage[most]{tcolorbox}
\usepackage{listings}

\definecolor{sigblue}{RGB}{229,232,252}

\newcolumntype{Y}{>{\raggedright\arraybackslash}X}

\newcommand{\wn}{Weibo Annotations}
\newcommand{\cn}{Community Notes}

\definecolor{PromptPurple}{HTML}{6F0074}

\lstdefinestyle{promptstyle}{
  basicstyle=\small\itshape,
  columns=fullflexible,
  keepspaces=true,
  breaklines=true,
  breakatwhitespace=false,
  showstringspaces=false,
  upquote=true,
  tabsize=2,
  aboveskip=0pt,
  belowskip=0pt
}

\newtcblisting{promptcard}[2][]{
  enhanced jigsaw,
  breakable,
  listing only,
  listing engine=listings,
  listing options={style=promptstyle},
  colback=white,
  colframe=PromptPurple,
  boxrule=0.8pt,
  arc=1.6mm,
  left=3.5mm,
  right=3.5mm,
  top=4.5mm,
  bottom=2.5mm,
  before skip=8pt,
  after skip=7pt,
  title={#2},
  fonttitle=\bfseries\footnotesize,
  coltitle=white,
  title after break={#2 (continued)},
  attach boxed title to top left={
    xshift=3mm,
    yshift=-2mm
  },
  boxed title style={
    colback=PromptPurple,
    colframe=PromptPurple,
    boxrule=0pt,
    arc=1mm,
    left=2.5mm,
    right=2.5mm,
    top=1mm,
    bottom=1mm
  },
  #1
}

\begin{document}

\title{When Platforms Pay Users to Correct Misinformation: Revenue Incentives and Governance Capacity in Weibo Annotations}
% \author{Anonymous Author(s)}

\author{Chichen Lin}
\authornote{These authors contributed equally to this work.}
\email{cuc\_lcc@cuc.edu.cn}
\affiliation{
  \institution{Communication University of China}
  \city{Beijing}
  \country{China}
}

\author{Huaijin Xie}
\authornotemark[1]
\email{xiehuaijin@cuc.edu.cn}
\affiliation{
  \institution{Communication University of China}
  \city{Beijing}
  \country{China}
}

\author{Wenhan Xie}
\email{wenhanxie@uchicago.edu}
\affiliation{
  \institution{University of Chicago}
  \city{Chicago}
  \state{Illinois}
  \country{United States}
}

\author{Weijian Fan}
\email{wjfan@cuc.edu.cn}
\affiliation{
  \institution{Communication University of China}
  \city{Beijing}
  \country{China}
}

\author{Xiao Han}
\email{hanxiao@cuc.edu.cn}
\affiliation{
  \institution{Communication University of China}
  \city{Beijing}
  \country{China}
}

\author{Yijie Jin}
\email{202312033012@mails.cuc.edu.cn}
\affiliation{
  \institution{Communication University of China}
  \city{Beijing}
  \country{China}
}

\author{Xiaojie Hao}
\email{202313093009@cuc.edu.cn}
\affiliation{
  \institution{Communication University of China}
  \city{Beijing}
  \country{China}
}

\author{Jiaying Wu}
\authornote{Corresponding author.}
\email{jiayingwu@u.nus.edu}
\affiliation{
  \institution{National University of Singapore}
  \city{Singapore}
  \country{Singapore}
}

\renewcommand{\shortauthors}{Lin et al.}

% tldr: Paying users mobilized corrective labor, but governance capacity remained constrained by participation, review, public reach, and contributor protection.

\begin{abstract}
Crowdsourced misinformation governance systems such as Community Notes and Weibo Annotations depend on sustained corrective labor. How monetary incentives shape participation and subsequent governance outcomes remains poorly understood. We conduct a mixed-methods study of Weibo Annotations, analyzing 102,994 posts and 431,407 comments from 2023--2026, alongside interviews with 15 contributors. We find concentrated corrective labor, limited observed display, and weak quality--engagement alignment. Policy introduction was associated with sharp increases in contributor participation and annotation production. Model-assisted quality followed its expected trajectory, evidence for additional display growth was inconclusive, and engagement per annotation fell below expectation. Public discussions and interviews reveal substantial verification demands, difficult-to-interpret review processes, uneven platform knowledge, and gaps in contributor protection. We identify four recurring bottlenecks in governance orchestration: participation, review, reach, and protection. Effective incentive design therefore depends on whether platforms can support these downstream processes as corrective labor expands.
\end{abstract}

\ccsdesc[500]{Human-centered computing~Empirical studies in collaborative and social computing}
\ccsdesc[300]{Human-centered computing~Social media}
\keywords{crowdsourced misinformation governance, corrective labor, governance orchestration, governance capacity, revenue incentives, Weibo Annotations}

\maketitle

\section{Introduction}
\begin{CJK*}{UTF8}{gbsn}
Social media spreads misinformation faster than professional fact-checkers and platform moderators can respond \cite{vosoughi2018spread,lazer2018science}. Crowdsourced misinformation governance systems such as X's \cn{}\footnote{\url{https://help.x.com/en/using-x/community-notes}} and Weibo's \wn{} (微博附注)\footnote{\url{https://www.weibo.com/ttarticle/p/show?id=2309404930349917667993}} respond to this gap by organizing ordinary users to identify questionable claims, evaluate evidence, and provide contextual corrections at platform scale \cite{proellochs2022birdwatch,saeed2022crowdsourced,lloyd2026beyond}. Research on \cn{} has examined contributor participation and selection, note evaluation, timeliness, and public effects \cite{chuai2024rollout,pilarski2024snoping,chuai2026request,chuai2026stability}. Much of this evidence comes from systems in which community ratings play a central role in selecting and surfacing corrections. \wn{} combines user-generated corrections with centralized platform review and platform-controlled display, creating a different governance structure for corrective participation. Recent work shows that Weibo annotations can improve misinformation correction \cite{dong2026platform}, while prior research documents Weibo's broader use of participatory moderation \cite{zhao2025governance}. How corrective labor is organized under this platform-mediated review and display structure remains less understood.
\end{CJK*}
Producing these corrections requires users to identify claims, gather evidence, assess sources, and write explanations. We refer to this effort as \textbf{corrective labor}. Platform incentives can shape this labor. Monetary rewards and visibility affect participation and content production on social media \cite{mason2009financial,sun2013ad,liu2021money,yi2025labor}, while misinformation research shows that incentives can influence information search, accuracy judgments, and fact-checking behavior \cite{panizza2022lateral,rathje2023accuracy,bodishtianu2025motivation}. However, expanding corrective labor addresses only production. Contributions must still be reviewed and made publicly reachable, while contributors need support to continue participating. This raises a broader question: \textbf{\textit{How do these platform processes respond when monetary incentives expand corrective labor?}}

We use \textbf{governance orchestration} to describe how platforms coordinate these processes as corrective labor expands. This includes sustaining participation, reviewing corrective contributions, providing public reach, and supporting contributors once their work becomes visible. We use \emph{governance capacity} to describe a platform's ability to support these functions at scale. Prior work identifies frictions across this process, including uneven verification effort \cite{chuai2026request,wack2026laziness}, unstable recognition \cite{chuai2026stability}, delayed public display \cite{chuai2024rollout}, and interaction costs surrounding visible corrections \cite{chuai2025outrage}. The central question is therefore whether governance capacity can keep pace with growth in corrective labor.

Weibo's May 2025 \textbf{Revenue Incentive Model}\footnote{\url{https://www.sina.cn/news/detail/5163349935653372.html}} provides a useful setting for this question. The policy pays eligible approved annotations CNY 5--50 and can redirect eligible advertising revenue from misleading posts to annotation contributors \cite{weibo2025incentive}. It increased the potential monetary return to corrective work, while review, display, and distribution remained governed by the broader platform process. This allows us to examine whether growth in participation and production was accompanied by changes in annotation quality, public display, and audience interaction, with public discussion and contributor interviews helping identify frictions as corrective labor expands.

The study asks:
\begin{description}
\item[RQ1:] How is corrective labor organized across Weibo Annotations?
\item[RQ2:] How did revenue incentives change corrective labor and downstream outcomes?
\item[RQ3:] What frictions shape participation, review, reach, and continued contribution?
\end{description}

We conduct a mixed-methods study of \wn{} spanning 2023--2026, analyzing 102,994 retrieved posts and 431,407 comments, alongside interviews with 15 contributors. Model-assisted annotation identifies substantive annotations, assesses annotation quality, and labels comment stance and toxicity. To examine policy-associated changes, we compare nine daily outcomes against Bayesian counterfactual trajectories around the May 2025 incentive policy. Public discussion and contributor interviews provide complementary evidence on how users experience verification, review, public reach, and social exposure.

The results reveal a clear asymmetry in the system's response to monetary incentives. Policy introduction was associated with sharp growth in contributor participation and annotation production, while model-assisted quality followed its expected trajectory, evidence for additional display growth was uncertain, and engagement per annotation fell below expectation. Public discussion and contributor interviews identify substantial verification demands, difficult-to-interpret review, uneven routes to public reach, and social risks surrounding visible corrections. We synthesize these constraints as \textbf{four recurring bottlenecks in governance orchestration: participation, review, reach, and protection.} This pattern highlights governance capacity as a key constraint on incentive-driven crowdsourced correction.

This work makes three contributions. Empirically, it provides a platform-scale mixed-methods account of a real-world revenue incentive in crowdsourced misinformation governance. Conceptually, it develops governance orchestration as an account of how participation, review, public reach, and contributor protection jointly shape governance capacity. For HCI design, it identifies concrete opportunities to support verification, review, distribution, and contributor protection.

\section{Related Work}

\subsection{Crowdsourced Misinformation Governance}

Crowdsourced misinformation governance involves ordinary users in identifying questionable claims, gathering evidence, writing corrections, and evaluating contributions from others. Such systems can provide fact-checking at scale and, under some conditions, produce judgments comparable to professional assessments \cite{allen2021scaling,proellochs2022birdwatch,saeed2022crowdsourced,drolsbach2024trust,borenstein-etal-2025-community}. Lloyd et al. conceptualize these platform-native systems as \emph{crowdsourced context systems} and organize their design into six dimensions: participation, inputs, curation, presentation, platform treatment, and transparency \cite{lloyd2026beyond}. Research on X's \cn{} further shows that system outcomes depend on how contributor attention is allocated \cite{pilarski2024snoping,pilarski2026supply,gong2026effects}, how contributions are evaluated and selected \cite{proellochs2022birdwatch,chuai2026stability,soham2025supernotes}, and whether approved notes become visible \cite{chuai2024rollout,slaughter2025community,chuai2026spread}. Participation and recognition can also become concentrated among highly active contributors, while consensus-based selection may disadvantage polarizing content \cite{nudo2026hyperactive,paul2026community}. 

Evidence on Chinese platforms remains more limited: prior work examines participatory moderation and contributor motivation on Weibo \cite{zhao2025governance}, while recent experimental research shows that \wn{} can improve audience responses to misinformation \cite{dong2026platform}. We extend this work by examining how corrective participation is shaped by platform review, public reach, and contributor support within \wn{}.

%%%%%%%%%%%%%%%%%%%%%%%%%%%%%%%%%%%%%%%%%%%%%%%%%%%%%%%%%%%%%%%%%%%%%%%%%%%%%%%%%%%%%%%%%%%%%%%%%%%%%%%%%%%%%%%%%

\subsection{Platform Incentives and Corrective Labor}

Crowdsourced misinformation governance depends on sustained user labor, which platforms shape through participation rules, evaluation mechanisms, visibility, and rewards \cite{lloyd2026beyond}. Research on online production shows that monetary rewards, reputation, revenue sharing, and platform monetization can influence participation and effort \cite{mason2009financial,sun2013ad,liu2021money,yi2025labor,eschebach2025google}. In misinformation contexts, monetary and accuracy incentives can similarly affect information search, truth discernment, and voluntary fact-checking \cite{panizza2022lateral,rathje2023accuracy,bodishtianu2025motivation}. These incentives operate within broader platform processes: requests can redirect corrective attention \cite{pilarski2026supply,gong2026effects}, successful publication can encourage subsequent contribution \cite{arjmandi2025threats}, and difficult claims can demand more verification effort and receive less corrective attention \cite{wack2026laziness,pilarski2024snoping}. Monetary incentives may therefore expand entry and production even as important participation constraints persist. Building on Lloyd et al.'s design space for crowdsourced context systems \cite{lloyd2026beyond}, we examine how participation, review, public reach, and contributor protection respond when monetary incentives expand corrective labor.

\subsection{Governance Orchestration}

Existing research identifies frictions across participation, review, public reach, and continued involvement. Participation can be concentrated among a small group of highly active contributors \cite{nudo2026hyperactive}; corrective attention can diverge from demand \cite{pilarski2026supply,gong2026effects}; review and selection depend on ratings, contributor composition, and consensus requirements \cite{chuai2026stability,paul2026community,soham2025supernotes}; and delays in public display can limit the effects of successful corrections \cite{chuai2024rollout,slaughter2025community,chuai2026spread}. Visible corrections can also reshape subsequent interactions and expose contributors to informational, emotional, and interpersonal costs \cite{chuai2025outrage,yoon2025collaborative}, while recognition, feedback, resource sharing, and social support can help sustain contribution \cite{arjmandi2025threats,yoon2025collaborative}. 

%%%%%%%%%%%%%%%%%%%%%%%%%%%%%%%%%%%%%%%%%%%%%%%%%%%%%%%%%%%%%%%%%%%%%%%%%%%%%%

This literature points to four recurring constraints: sustaining participation, making review legible to contributors, securing public reach for corrective work, and protecting contributors after visibility. How these constraints interact when incentives rapidly expand corrective labor remains less understood.

%%%%%%%%%%%%%%%%%%%%%%%%%%%%%%%%%%%%%%%%%%%%%%%%%%%%%%%%%%%%%%%%%%%%%%%%
\FloatBarrier

%%%%%%%%%%%%%%%%%%%%%%%%%%%%%%%%%%%%%%%%%%%%%%%%%%%%%%%%%%%%%%%%%%%%%%%%%%%%%%%%%%%%%%%%%%%%%%%%%%%%%%%%%%%%%%%%%
\begin{figure*}[t]
  \centering
  \includegraphics[width=\textwidth]{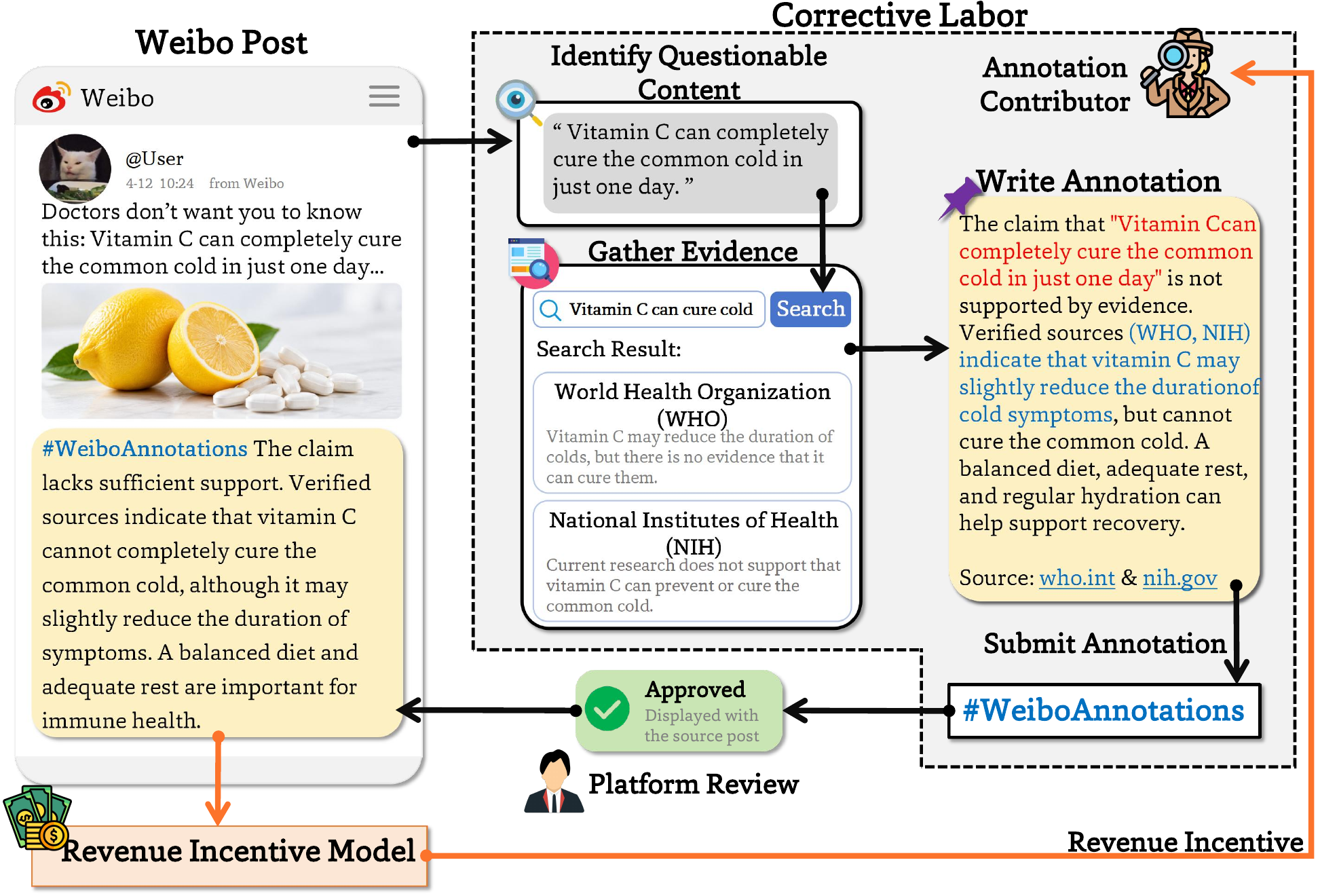}
  \caption{Workflow of Weibo Annotations and the intervention point of the Revenue Incentive Model, reconstructed from Weibo's public announcement \cite{weibo2025incentive}. Contributors identify questionable content, gather evidence, write and submit annotations, which are then reviewed by the platform and, if approved, displayed alongside the source post. The Revenue Incentive Model introduces monetary rewards for eligible approved annotations and may redirect eligible advertising revenue associated with misleading posts to annotation contributors. The orange path highlights where the incentive enters this process: it provides monetary returns to contributors after approval. The complete English translation of the policy appears in Appendix~\ref{app:policy-translation}.}
  \label{fig:policy-context}
\end{figure*}
\vspace{-6pt}
\section{Weibo Annotations and the Revenue Incentive Model}
\label{sec:weibo-context}

\wn{} is a crowdsourced governance feature that allows users to add corrections, contextual information, supporting sources, or affected-party responses to public posts \cite{weibo2026esg,li2026weibo}. As shown in Figure~\ref{fig:policy-context}, contributors identify potentially questionable content, gather evidence, and submit annotations for centralized platform review. Approved annotations may then be displayed alongside the source post and receive public engagement through likes, comments, and reposts. The workflow therefore links corrective production with platform review, public display, and audience interaction.

Weibo announced the \emph{Revenue Incentive Model} on May 6, 2025 \cite{weibo2025incentive}. Beginning May 7, eligible approved annotations could receive CNY 5--50, while eligible advertising revenue associated with misleading posts could also be redirected to annotation contributors. Only the first approved annotation on a source post is rewarded, and each account can receive incentives for at most three annotations per day \cite{weibo2025incentive}. Figure~\ref{fig:policy-context} situates these monetary incentives within the broader annotation workflow.

The policy introduced monetary rewards for eligible approved contributions, while review and public display remained governed separately by the platform. This separation provides a useful setting for examining whether stronger incentives are accompanied by changes in corrective participation, production, public display, and audience interaction.

\section{Methods}
\label{sec:methods}

\begin{figure*}[t]
    \centering
    \includegraphics[width=1\textwidth]{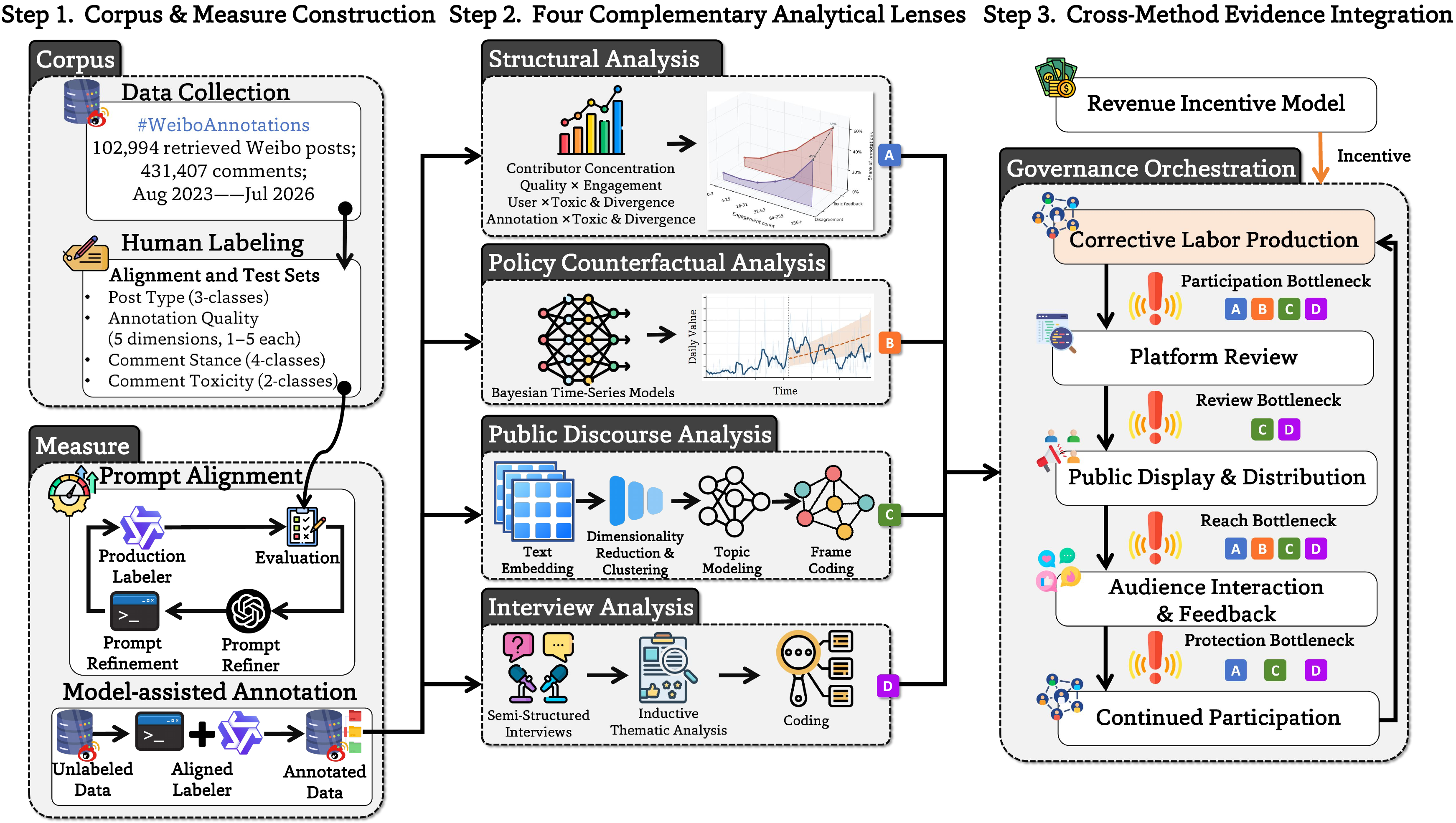}
    \caption{Mixed-methods research design and cross-method evidence integration. Step 1 constructs the corpus and model-assisted measures using human-labeled alignment and held-out test sets. Step 2 applies four complementary analytical lenses: (A) structural analysis, (B) policy counterfactual analysis, (C) public discourse analysis, and (D) interview analysis. Step 3 integrates evidence across the governance orchestration process, covering corrective labor production, platform review, public display and distribution, audience interaction and feedback, and continued participation. The A--D tags indicate which analytical lenses inform the participation, review, reach, and protection bottlenecks.}
    \label{fig:method-workflow}
\end{figure*}

We use a mixed-methods design to examine the structure of corrective labor, policy-associated changes, and user experiences across the governance process. Figure~\ref{fig:method-workflow} summarizes the study in three steps. Step 1 constructs the corpus and model-assisted measures using human-labeled alignment and held-out test sets. Step 2 applies four complementary analyses: structural analysis for RQ1, policy counterfactual analysis for RQ2, and public discourse and interview analyses for RQ3. Step 3 integrates evidence across these analyses to identify recurring constraints in governance orchestration.
\paragraph{\textbf{Ethical considerations.}} The study received approval from the relevant institutional ethics review committee. Interview participants provided informed consent prior to participation. The social media analysis used publicly accessible Weibo content; additional procedures concerning data handling, participant privacy, and interview recruitment are described in Appendix \ref{app:interviews}.

%%%%%%%%%%%%%%%%%%%%%%%%%%%%%%%%%%%%%%%%%%%%%%%%%%%%%%%%%%%%%%%%%%%%%%%%%%%%%%
\vspace{-10pt}

\subsection{Step 1: Corpus and Measure Construction}
\label{sec:data}
\label{sec:model-assisted}

\paragraph{\textbf{Corpus construction.}}
\begin{CJK*}{UTF8}{gbsn}
We collected public Weibo posts associated with the \wn{} hashtag (\#微博附注) and related annotation markers, together with available metadata, linked source posts, and public comments. The corpus covers August 17, 2023 to July 14, 2026 and contains 102,994 retrieved posts and 431,407 analyzable comments.
\end{CJK*}

Retrieved posts are classified as \emph{substantive annotations}, \emph{mechanism-discussion candidates}, or irrelevant posts. Substantive annotations correct, contextualize, support, or otherwise respond to a source post through \wn{}, while mechanism-discussion candidates concern the annotation system itself. This yields 42,853 substantive annotations, 1,100 mechanism-discussion candidates, and 59,041 irrelevant posts; the substantive annotations are associated with 141,736 analyzable comments. Appendix~\ref{app:corpus} reports corpus construction details.

\paragraph{\textbf{Human labeling and model-assisted annotation.}}
We constructed human-labeled alignment and held-out test sets for four tasks: post type, annotation quality, comment stance, and comment toxicity. Annotation quality is assessed along five dimensions: relevance, evidence, objectivity, helpfulness, and fluency, each scored on a five-point Likert scale. The primary quality measure is the unweighted mean of these five scores.

Qwen3-4B serves as the production model for model-assisted annotation \cite{yang2025qwen3}. Inspired by AlphaEvolve and PromptAgent \cite{alphaevolve,promptagent}, we iteratively refined task prompts on the alignment sets through evaluation and error analysis, then froze them before held-out testing and full-corpus inference. Across the four tasks, test-set accuracy ranged from .785 to .895. For annotation quality, formulated as a five-level ordinal regression task, exact-match accuracy was .833, weighted $\kappa=.662$, and MAE $=.325$. Appendix~\ref{app:audits} reports the human-labeling protocol, prompt alignment procedure, held-out performance, and error analysis.

\paragraph{\textbf{Measures and observable scope.}}
We characterize corrective labor through annotation production, contributor activity, and model-assisted annotation quality. Downstream outcomes capture collection-time display, recorded engagement, and audience response. Engagement is defined as $\log(1+\text{likes}+\text{comments}+\text{reposts})$, while audience response is characterized through comment volume, disagreement, and toxicity. Full definitions and aggregation rules appear in Appendices~\ref{app:corpus} and~\ref{app:audits}.

Public traces omit complete review histories, rankings, impressions, and payment records. Collection-time display and recorded engagement therefore serve as observable indicators of public reach. Platform approval histories, actual exposure, and persuasive effects remain outside the observable scope.

\subsection{Step 2: Four Complementary Analytical Lenses}
% \subsection{Step 2: Mixed-Methods Analysis}
\label{sec:analysis}

\paragraph{\textbf{A. Structural Analysis (RQ1).}} To characterize how corrective labor is organized and connected to observable downstream outcomes, we examine contributor concentration, collection-time display, the association between annotation quality and engagement, and audience feedback across contributor-activity and engagement strata. The analysis uses concentration measures, Spearman rank associations, and descriptive bootstrap summaries \cite{efron1979bootstrap}.

\paragraph{\textbf{B. Policy Counterfactual Analysis (RQ2).}} To assess how corrective labor and downstream outcomes changed around the incentive policy, we examine nine daily outcomes over symmetric 90-day windows around the May 7, 2025 Revenue Incentive Model. Bayesian time-series models fitted to the pre-policy period project no-policy counterfactual trajectories using linear trends, weekday effects, and AR(1) residuals \cite{brodersen2015causalimpact}. Daily display and engagement are indexed by annotation posting date and use collection-time observations. Robustness checks use alternative temporal windows, segmented interrupted time-series models \cite{bernal2017its}, placebo interventions, and rolling forecasts. Appendix~\ref{app:counterfactual-model} reports the full specification and diagnostics.

\paragraph{\textbf{C. Public Discourse Analysis (RQ3).}} To capture how users publicly interpret and evaluate the annotation mechanism, three researchers screened 1,100 mechanism-discussion candidates and retained 291 confirmed discussions. Semantic clustering surfaced recurrent topics, which the researchers labeled and grouped into broader public-discussion frames. The resulting frame definitions were then applied at the document level to all confirmed discussions, assigning each post a primary frame for a descriptive pre/post comparison around the policy launch. Appendix~\ref{app:frames} reports the screening, clustering, frame coding, and comparison procedures.

\paragraph{\textbf{D. Interview Analysis (RQ3).}} To understand how contributors experience the annotation process in practice, we conducted 15 semi-structured interviews lasting 42--100 minutes in November and December 2025. Purposive sampling captured variation in contribution frequency, gender, and verification status. 
Inductive thematic analysis identified recurring experiences around verification, review, public reach, interaction, and continued participation. Appendix~\ref{app:interviews} reports the recruitment, interview protocol, and analytic procedure.

%%%%%%%%%%%%%%%%%%%%%%%%%%%%%%%%%%%%%%%%%%%%%%%%%%%%%%%%%%%%%%%%%%%%%%%%%%%%%%
\begin{figure*}[t]
\centering
\includegraphics[width=0.95\textwidth]{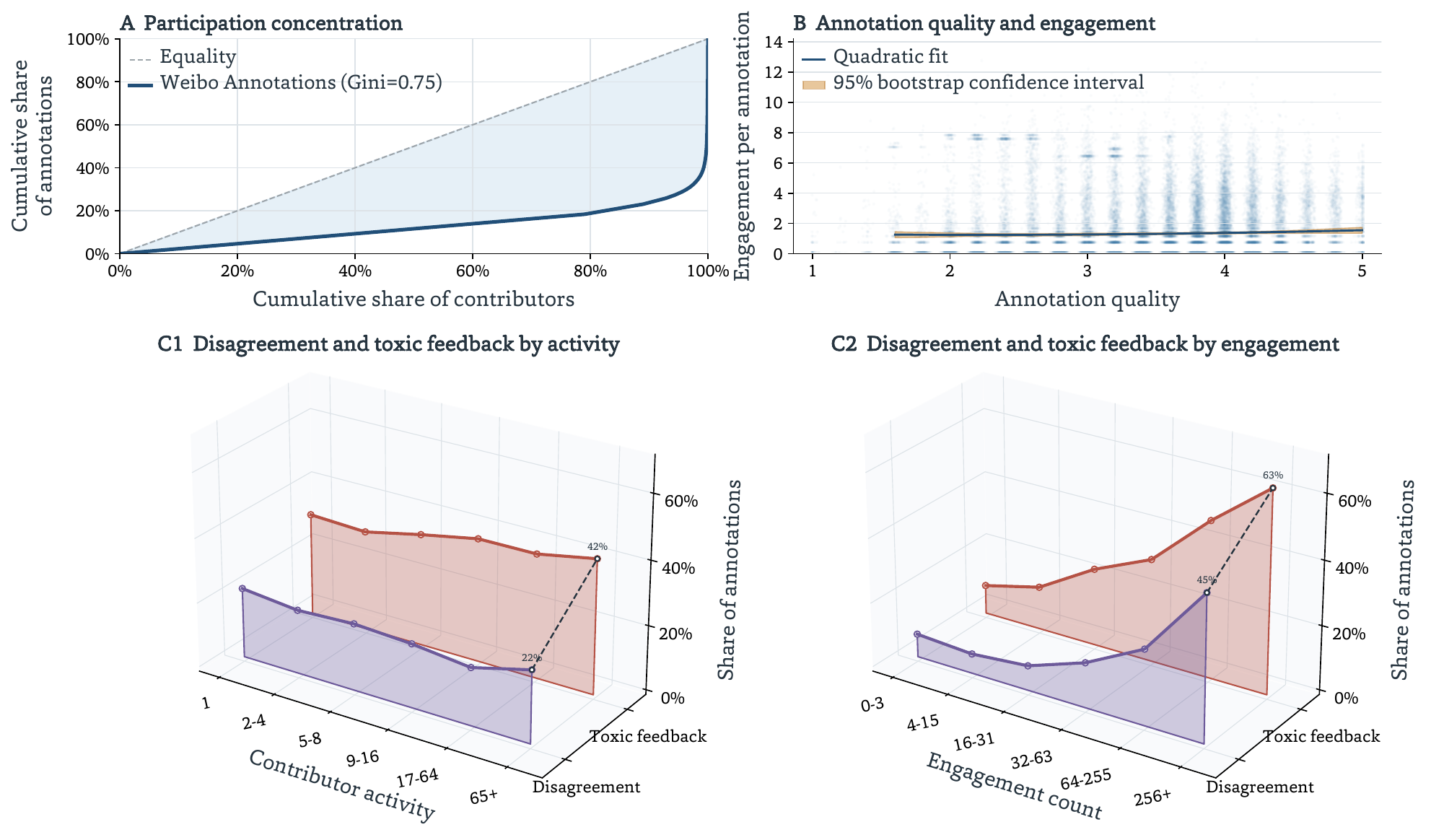}
  \caption{\textbf{Structural patterns in 42,853 substantive annotations.} (A) Production concentration across 9,917 contributors. (B) Five-point model-assisted annotation quality and engagement per annotation; the line is a descriptive quadratic fit. (C1--C2) Disagreement and toxic feedback among 6,735 annotations with at least two comments, stratified by contributor activity and annotation engagement. Measure definitions are provided in Appendix~\ref{sec:measures}.}
\Description{Four panels show a strongly bowed Lorenz curve for annotation production, a diffuse scatterplot with a nearly flat relationship between model-assisted annotation quality and engagement, and the shares of eligible annotations with disagreement or toxic feedback across contributor-activity and engagement strata.}
\label{fig:rq1-structure}
\end{figure*}
%%%%%%%%%%%%%%%%%%%%%%%%%%%%%%%%%%%%%%%%%%%%%%%%%%%%%%%%%%%%%%%%%%%%%%%%%%%%%%

% \subsection{Step 3: Findings Integration}
\subsection{Step 3: Cross-Method Evidence Integration}
\label{sec:integration}

After completing the four analyses separately, we compare their findings across the governance process shown in Figure~\ref{fig:method-workflow}. Behavioral patterns, policy-associated changes, public discourse, and contributor accounts are used to identify recurring constraints that appear across methods or stages. This synthesis organizes the evidence into four bottleneck classes in governance orchestration: participation, review, reach, and protection. The A--D tags in Figure~\ref{fig:method-workflow} indicate which analytical lenses contribute evidence to each class.

%%%%%%%%%%%%%%%%%%%%%%%%%%%%%%%%%%%%%%%%%%%%%%%%%%%%%%%%%%%%%%%%%%%%%%%%%%%%%%
%%%%%%%%%%%%%%%%%%%%%%%%%%%%%%%%%%%%%%%%%%%%%%%%%%%%%%%%%%%%%%%%%%%%%%%%%%%%%%
\begin{table}[htbp]
\centering
\small
\caption{Observable states of substantive annotations at collection time ($N=42{,}853$).}
\label{tab:rq1-surfacing}
\begin{tabularx}{\columnwidth}{@{}Yrr@{}}
\toprule
\textbf{Observable state} & \textbf{$N$} & \textbf{Share} \\
\midrule
Linked and displayed & 3,371 & 7.9\% \\
Linked, undisplayed & 23,375 & 54.5\% \\
Linked, source inaccessible & 14,857 & 34.7\% \\
No source link & 1,244 & 2.9\% \\
Display status unresolved & 6 & $<0.1\%$ \\
\bottomrule
\end{tabularx}
\end{table}
%%%%%%%%%%%%%%%%%%%%%%%%%%%%%%%%%%%%%%%%%%%%%%%%%%%%%%%%%%%%%%%%%%%%%%%%%%%%%%

\section{Results}
\label{sec:results}

The three RQs build on one another. RQ1 establishes the structural patterns of corrective labor and observable public outcomes(Section~\ref{sec:rq1}); RQ2 examines policy-associated changes in these outcomes (Section~\ref{sec:rq2}); and RQ3 identifies process-level frictions that help contextualize the observed divergence (Section~\ref{sec:rq3}).

\subsection{RQ1: Concentrated Corrective Labor, Limited Observed Display, and Weak Quality--Engagement Link}
\label{sec:rq1}

Across the full observation period, 9,917 accounts produced 42,853 substantive annotations. Four structural patterns stand out: corrective production was highly concentrated, few annotations were observed as displayed at collection time, model-assisted quality was weakly associated with recorded engagement, and higher-engagement annotations received more mixed and toxic feedback.

\paragraph{\textbf{Corrective Production Was Highly Concentrated}}

Corrective production had a Gini coefficient of .749. The most active 10\% of contributors produced 76.2\% of substantive annotations, while 78.7\% of contributors appeared only once in the corpus. This concentration was greater than reported for Community Notes, where the top 10\% produced 58\% of notes and the Gini coefficient was .68~\cite{razuvayevskaya2025timeliness}. Observed production on \wn{} was therefore dominated by a relatively small active core.

\paragraph{\textbf{Only a Small Share of Annotations Was Observed as Publicly Displayed}}
At collection time, 3,371 substantive annotations (7.9\%) were visibly attached to an accessible source post (Table~\ref{tab:rq1-surfacing}). Among annotations with an accessible source post and resolvable display state, 12.6\% were displayed. Display is measured at collection time and captures whether an annotation was visibly attached to its source post; prior display states and review outcomes remain unobserved.

\paragraph{\textbf{Annotation Quality Was Weakly Related to Audience Engagement}}

Across substantive annotations, model-assisted quality had only a small rank association with recorded engagement (Spearman $\rho=.068$), and 52.3\% received no recorded likes, comments, or reposts. The fitted relationship in Figure~\ref{fig:rq1-structure} was similarly shallow across most of the quality range. Higher assessed quality therefore corresponded only weakly to greater recorded interaction.

\paragraph{\textbf{Greater Engagement Coincided with More Mixed and Toxic Feedback}}

Audience feedback varied more strongly with engagement than with contributor activity. Among the 6,735 annotations with at least two comments, the share containing both supportive and questioning or opposing comments ranged only from 17.8\% to 22.5\% across contributor-activity strata. Mean within-annotation toxicity increased from 31.8\% for one-time contributors to 41.8\% for contributors with 65 or more annotations.

The pattern was substantially stronger across engagement strata. The share of annotations containing both supportive and questioning or opposing comments rose from 7.2\% in the 0--3 engagement group to 45.2\% in the 256-or-more group. Mean within-annotation toxicity increased from 9.0\% to 62.9\% across the same strata (Figure~\ref{fig:rq1-structure}). Because observing both stance types becomes more likely as comment volume increases, the mixed-response pattern is interpreted descriptively. Higher recorded engagement nevertheless coincided with a substantially more toxic feedback environment, echoing prior work on contestation in crowdsourced fact-checking~\cite{razuvayevskaya2025timeliness}.

%%%%%%%%%%%%%%%%%%%%%%%%%%%%%%%%%%%%%%%%%%%%%%%%%%%%%%%%%%%%%%%%%%%%%%%%%%%%%%
\begin{figure*}[t]
\centering
\includegraphics[width=0.95\textwidth]{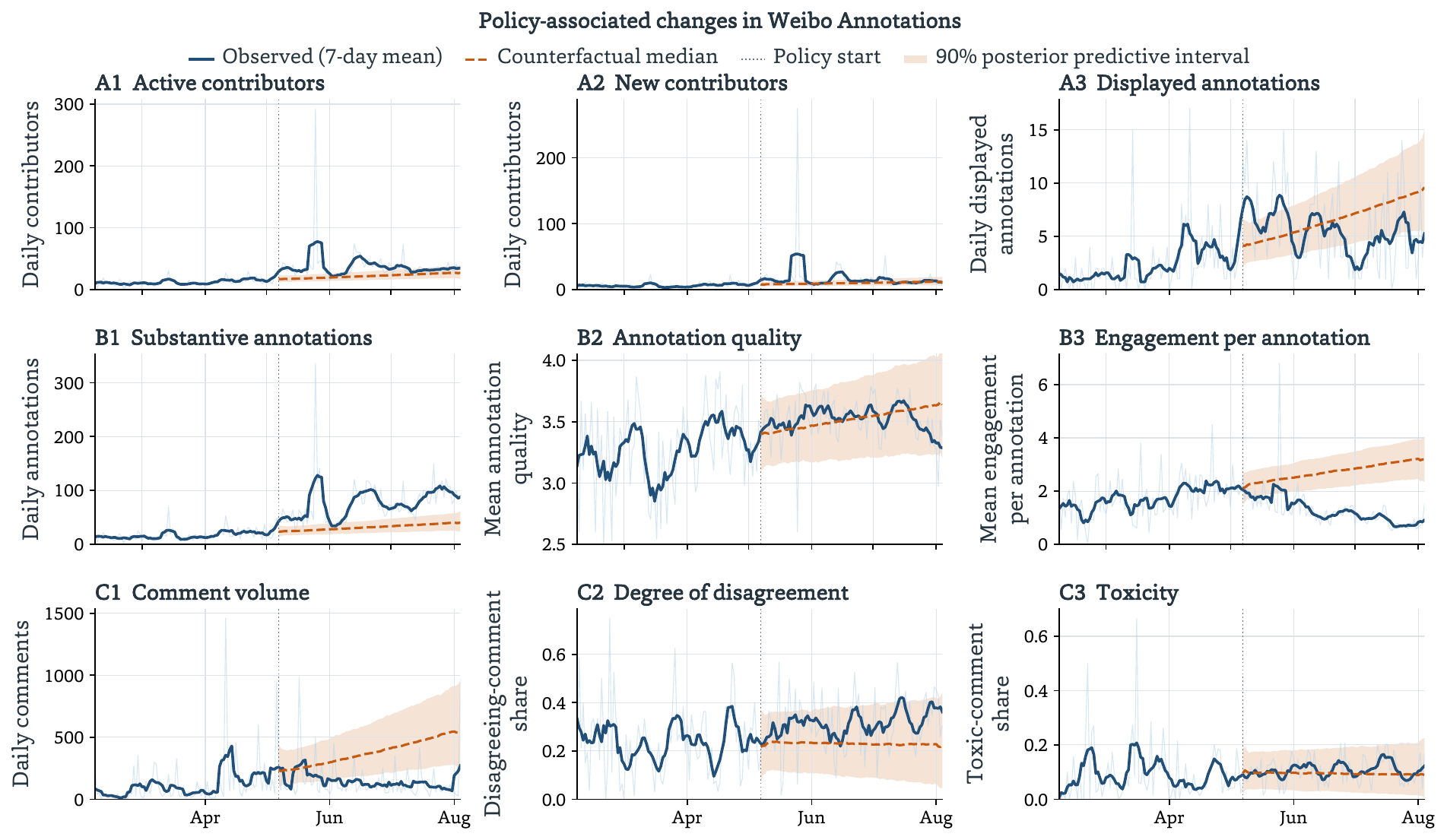}
\caption{\textbf{Policy-associated changes in the symmetric 90-day window around May 7, 2025.} Pale lines are daily observations, solid lines centered seven-day means, dashed lines no-policy counterfactual posterior medians, and shaded bands 90\% posterior predictive intervals. Definitions and scales for the nine outcomes are provided in Appendix~\ref{sec:measures}.}
\Description{Nine time-series panels compare observed outcomes with estimated no-policy trajectories. Active contributors, new contributors, and substantive annotations increase following the policy launch. Displayed annotations remain close to the counterfactual, annotation quality follows its expected trajectory, and engagement per annotation falls below expectation. Comment volume stays below the counterfactual trajectory, while degree of disagreement and toxicity show no reliable divergence.}
\label{fig:policy-dynamics}
\end{figure*}
%%%%%%%%%%%%%%%%%%%%%%%%%%%%%%%%%%%%%%%%%%%%%%%%%%%%%%%%%%%%%%%%%%%%%%%%%%%%%%

\begin{table*}[htbp]
\centering
\footnotesize
\caption{\textbf{Observed changes and policy-associated gaps during the
90-day policy window.}
Gaps compare observed post-policy values with no-policy counterfactuals;
brackets report 90\% Credible Intervals (CrI), and bold indicates intervals excluding zero.}
\label{tab:counterfactual}

\setlength{\tabcolsep}{5pt}
\renewcommand{\arraystretch}{0.92}

\begin{tabularx}{\textwidth}{
@{}>{\raggedright\arraybackslash}X
r
r
r
>{\raggedleft\arraybackslash}p{0.29\textwidth}
@{}}
\toprule
\textbf{Measure}
& \textbf{Pre}
& \textbf{Post}
& \textbf{Counterfactual}
& \textbf{Policy gap [90\% CrI]} \\
\midrule

\multicolumn{5}{@{}l}{\textbf{User level}} \\[-2pt]
Active contributors
& 12.36 & 37.69 & 22.06
& \textbf{$+70.9\%$ [$+34.9\%, +118.2\%$]} \\
New contributors
& 6.24 & 16.71 & 9.97
& \textbf{$+67.6\%$ [$+21.0\%, +142.7\%$]} \\

\midrule
\multicolumn{5}{@{}l}{\textbf{Annotation level}} \\[-2pt]
Substantive annotations
& 17.01 & 76.58 & 31.97
& \textbf{$+139.5\%$ [$+84.5\%, +220.4\%$]} \\
Annotation quality
& 3.27 & 3.52 & 3.52
& $-0.001$ [$-0.239, +0.240$] \\
Displayed annotations
& 2.32 & 5.54 & 6.64
& $-16.5\%$ [$-38.0\%, +18.6\%$] \\

\midrule
\multicolumn{5}{@{}l}{\textbf{Response level}} \\[-2pt]
Engagement per annotation
& 1.73 & 1.28 & 2.73
& \textbf{$-1.46$ [$-1.90, -1.00$]} \\
Comment volume
& 124.28 & 146.33 & 382.48
& \textbf{$-61.7\%$ [$-73.7\%, -39.3\%$]} \\
Degree of disagreement
& .239 & .314 & .231
& $+.083$ [$-.031, +.186$] \\
Toxicity
& .092 & .106 & .096
& $+.011$ [$-.058, +.060$] \\

\bottomrule
\end{tabularx}
\end{table*}
%%%%%%%%%%%%%%%%%%%%%%%%%%%%%%%%%%%%%%%%%%%%%%%%%%%%%%%%%%%%%%%%%%%%%%%%%%%%%%

% \subsection{RQ2: Revenue Incentives Produced Uneven Changes Across the Governance Process}
\subsection{RQ2: Revenue Incentives Expanded Corrective Labor, with Uneven Downstream Change}
\label{sec:rq2}

Figure~\ref{fig:policy-dynamics} and Table~\ref{tab:counterfactual} show the strongest policy-associated departures in contributor participation and annotation production. Model-assisted quality remained close to its expected trajectory, the 90-day display estimate was uncertain, and engagement per annotation fell below expectation. Sensitivity analyses reproduced the main labor-mobilization pattern, while placebo diagnostics identified production, active contributors, and declining engagement as the clearest departures (Appendix~\ref{app:counterfactual-model}).

\paragraph{\textbf{User Level: Contributor Entry and Activity Increased}} Active contributors were 70.9\% above the estimated no-policy trajectory [34.9\%, 118.2\%], while new contributors were 67.6\% above expectation [21.0\%, 142.7\%]. Both increases were strongest soon after policy introduction and attenuated later in the 90-day post-policy window.

% \paragraph{\textbf{Annotation Level: Production Increased; Quality and Display Changed Less}}
\paragraph{\textbf{Annotation Level: Production Increased, While Quality and Display Showed Limited Change}} 
Substantive annotations were 139.5\% above the counterfactual trajectory [84.5\%, 220.4\%], the largest relative increase among the count outcomes. Model-assisted annotation quality remained close to expectation, with an estimated gap of $-0.0005$ [$-0.2385$, $+0.2395$] on the five-point scale.

Displayed annotations rose shortly after policy introduction and then declined. Across the full 90-day window, the estimated gap was $-16.5\%$ [$-38.0\%$, $+18.6\%$], with the interval spanning zero. The data therefore provide no clear evidence of additional display growth over the full post-policy period.

% \paragraph{\textbf{Interaction Level: Engagement per Annotation Declined}}
\paragraph{\textbf{Interaction Level: Engagement per Annotation Fell Below Expectation}} 
Mean log-transformed engagement per annotation was $1.456$ below its counterfactual trajectory [$-1.903$, $-1.003$], with the gap developing gradually over the post-policy period.

Daily comment volume increased in raw mean from 124.3 before the policy to 146.3 afterward, but remained 61.7\% below the estimated counterfactual [$-73.7\%$, $-39.3\%$]. Placebo diagnostics showed greater pre-policy variability for this outcome, so we interpret this estimate cautiously. Comment composition showed no clear departure: disagreement was $+0.0829$ [$-0.0310$, $+0.1861$] above expectation and toxicity was $+0.0105$ [$-0.0577$, $+0.0603$].

RQ2 therefore reveals a marked divergence between labor mobilization and downstream trajectories. Contributor participation and corrective production increased sharply around policy introduction, while display showed no clear additional growth and engagement per annotation fell below expectation. RQ3 next examines the public discussions and contributor experiences that provide process-level context for this divergence.

%%%%%%%%%%%%%%%%%%%%%%%%%%%%%%%%%%%%%%%%%%%%%%%%%%%%%%%%%%%%%%%%%%%%%%%%%%%%%%
\begin{figure*}[t]
\centering
\includegraphics[width=0.98\textwidth]{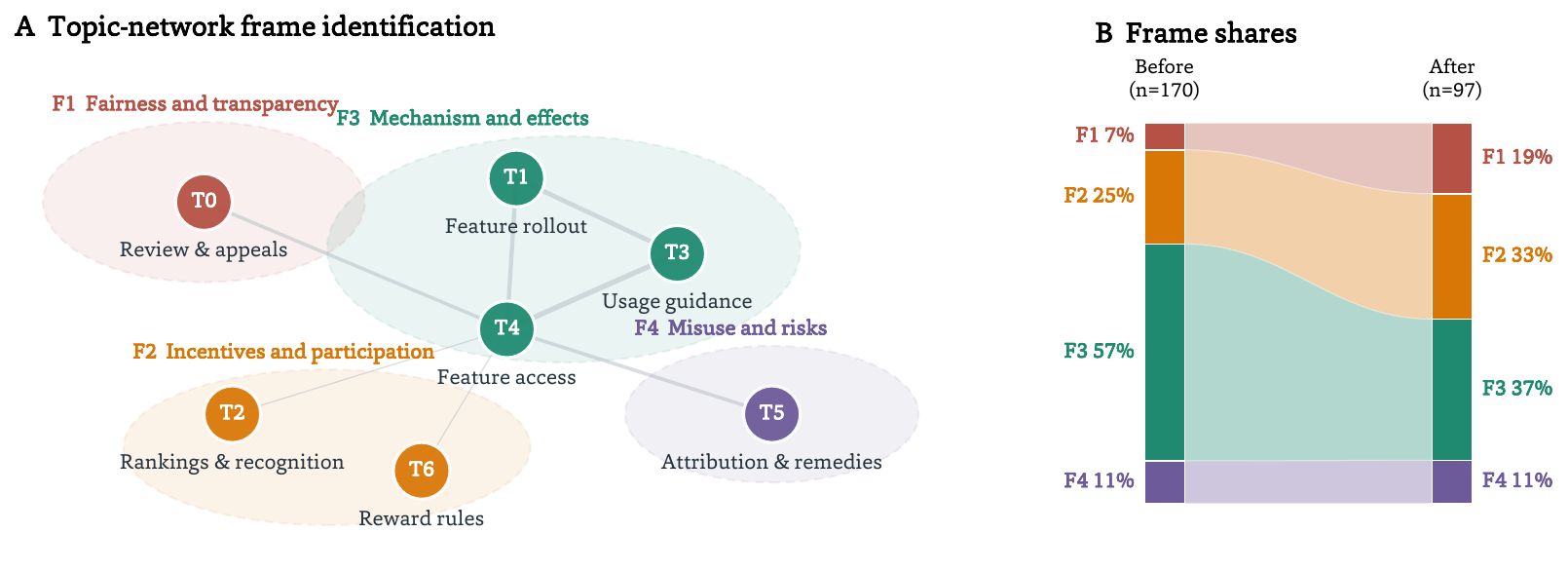}
\caption{\textbf{Public discussion of the Weibo Annotations mechanism.} (A) Seven topics manually grouped into four frames, with links indicating semantic similarity. (B) Descriptive frame shares in the pre- and post-policy windows.}
\label{fig:public-discourse}
\end{figure*}
%%%%%%%%%%%%%%%%%%%%%%%%%%%%%%%%%%%%%%%%%%%%%%%%%%%%%%%%%%%%%%%%%%%%%%%%%%%%%%

\subsection{RQ3: Recurring Frictions in Verification, Review, Reach, and Protection}
\label{sec:rq3}

Public discussion and contributor interviews identified recurring frictions in verification, review, public reach, and social exposure. Figure~\ref{fig:public-discourse}A organizes public discussion into four researcher-coded frames: F1 \emph{Fairness and transparency}, F2 \emph{Incentives and participation}, F3 \emph{Mechanism and effects}, and F4 \emph{Misuse and risks}, together with their associated topics. Figure~\ref{fig:public-discourse}B provides a descriptive comparison of frame prominence around the incentive policy. Descriptively, post-policy discussion shifted away from \emph{mechanism and effects} (57\% to 37\%) and toward \emph{fairness and transparency} (7\% to 19\%) and \emph{incentives and participation} (25\% to 33\%), while \emph{misuse and risks} remained stable at 11\%. This shift is consistent with review outcomes and reward allocation becoming more salient once approval became linked to monetary rewards. The F1--F4 labels indicate which public-discussion frames contribute evidence to each friction, while the frictions themselves synthesize evidence across public discussion and contributor interviews.

\paragraph{\textbf{Verification Demands Raised Participation Costs}} 

Public discussion in F2 \emph{Incentives and participation} showed that monetary rewards had become salient in annotation work. Users discussed recognition, competition for annotations, and perceived advantages from low-effort or copied contributions. One post-policy user described rewards as difficult to earn because annotation required substantial verification skill and responsibility. These discussions highlighted tension between a common reward structure and widely varying verification demands.

Interviews made this variation concrete. Rewards entered existing motivations, including maintaining a recognizable public persona, helping others judge disputed information, and earning ``a little pocket money.'' They could also influence which posts contributors chose to annotate. Verification demands, however, varied substantially across cases. Contributors described searching for authoritative sources, comparing evidence, and carefully formulating claims. A2 estimated that annotating another user's post took ``at least three to five times as long as writing an ordinary post,'' with conflicting authoritative sources creating additional judgment demands. The same reward structure therefore applied to corrective work requiring very different levels of time, knowledge, and effort.

\paragraph{\textbf{Review Outcomes Were Hard to Predict and Learn From}}

F1 \emph{Fairness and transparency} centered heavily on review and appeals. Users described similar submissions receiving different outcomes, approved annotations being reconsidered, and cases they perceived as inconsistently reviewed. Some users further perceived bias in platform review, suspecting that particular accounts or topics received different levels of scrutiny. These accounts do not establish differential treatment, but they show how opaque outcomes can generate person- and topic-specific theories of platform review.

Interviews showed how this uncertainty shaped subsequent contribution. A1 reported that some annotations passed within ``four or five minutes,'' whereas others remained under review much longer. A2 compared submission to ``buying a lottery ticket.'' Contributors considered evidence quality, wording, duplication, timing, and topic sensitivity as possible explanations for unsuccessful submissions, often without clear feedback. Tracking was also difficult: A6 described having ``only a submission interface and no dedicated archive.'' A14, who had access to dedicated customer support, changed his future case selection after receiving an explanation for a failed annotation. Review outcomes therefore provided weak and unevenly accessible signals for learning where future effort might be worthwhile.

\paragraph{\textbf{Public Reach Was Shaped by Platform Literacy and Social Resources}}

Public discussion in F3 \emph{Mechanism and effects} showed that effective use of \wn{} required practical knowledge extending beyond annotation writing itself. Users described learning through repeated use, navigating review and appeals, and working around constraints on presenting multimedia evidence and sources. Other discussions highlighted platform contingencies that could disrupt circulation, such as source-post deletion removing the immediate context for a correction. These accounts suggest a gap between formal access to the annotation mechanism and effective access to its audience.

Interviews revealed a parallel dependence on social resources. A11 described relying on an existing fan community for early likes and comments that helped an annotation gain traction. Public reach could therefore depend partly on platform literacy and pre-existing social ties that were unevenly distributed across contributors. This made visibility an additional resource that contributors had to secure after producing a correction.

\paragraph{\textbf{Visible Corrections Exposed Users to Conflict and Harassment}}

F4 \emph{Misuse and risks} captured concerns that visible annotations could become entangled with interpersonal and group conflict. Users described repeated or selective annotation as harassment or as a ``weapon'' against opposing camps. Corrective annotations could also trigger retaliation when they affected others' visibility or perceived interests, including insults, blocking, and other forms of pressure. Some post-policy discussions further raised concerns that rewards could encourage repetitive or misleading annotations, adding another source of conflict around the mechanism.

Contributors described similar risks after annotations gained visibility. A3 reported receiving thousands of hostile comments:

\begin{quote}
\textbf{``I was cursed at in thousands of comments. Some people first asked politely for the annotation to be removed; when I did not respond, they attacked and blocked me.''}

\hfill ——A3
\end{quote}

A7 also observed multiple accounts posting highly similar comments calling on the platform to punish an annotated user or members of an opposing camp. A7's account illustrates how annotation disputes could become entangled with existing group conflicts and collective pressure. Contributors and annotated users could therefore both become targets of harassment surrounding visible corrections.

Across RQ3, the evidence reveals four recurring frictions: uneven verification costs, opaque and weakly learnable review, unequal access to public reach, and social exposure after visibility. These frictions provide process-level context for the divergence observed in RQ1 and RQ2. Section~\ref{sec:discussion} develops them as participation, review, reach, and protection bottlenecks in governance orchestration.

%%%%%%%%%%%%%%%%%%%%%%%%%%%%%%%%%%%%%%%%%%%%%%%%%%%%%%%%%%%%%%%%%%%%%%%%%%%%%%%%%%%%%%%%%%%%%%%%%%%%

\section{Discussion}
\label{sec:discussion}

Our findings reveal asymmetric scaling across the governance process. Around the Revenue Incentive Model, contributor entry and annotation production rose sharply, while model-assisted quality followed its expected trajectory, display showed no clear additional growth over the full policy window, and engagement per annotation fell below expectation. Public discussion and interviews locate recurring constraints around verification, review, public reach, and social exposure. We use \emph{governance orchestration} to describe the coordination required to sustain participation, process corrective work, provide public reach, and support users as corrective labor expands. The four bottlenecks below identify distinct capacity constraints within this process.

% \subsection{Participation Bottleneck: Entry and Persistence}
\subsection{Participation Bottleneck: Entry and Continued Contribution}

A \emph{participation bottleneck} arises when increased entry is difficult to convert into continued corrective contribution. In \wn{}, production remained concentrated among a small contributor core, while the sharp increase in submission participation after policy introduction attenuated over the post-policy period. Public discussion and interviews further show that annotation can require substantial search, source evaluation, and judgment, especially when evidence is difficult to locate or conflicting.

Monetary rewards increase the potential benefit of contributing, while verification demands remain uneven across cases. Contributors still need sufficient time, knowledge, and confidence to produce defensible corrections, consistent with prior work showing that verification costs shape corrective attention \cite{wack2026laziness,pilarski2024snoping}. The participation bottleneck therefore captures the difficulty of sustaining corrective labor once the initial incentive to enter has been provided.

% \subsection{Review Bottleneck: Capacity and Feedback}
\subsection{Review Bottleneck: Legibility and Feedback}

A \emph{review bottleneck} arises when contributors have difficulty predicting, tracking, or learning from review outcomes. Review determines whether corrective work is recognized by the platform, becomes eligible for reward, and can proceed toward public display. Contributors described variable review times and outcomes, limited explanations, difficulty tracking prior submissions, and unequal access to support.

Review also provides information that shapes future contribution. Prior work shows that users can develop practical understandings of opaque platform systems through observable outcomes \cite{eslami2016folk,devito2018folk,mayworm2024moderation}; our findings show a similar process around annotation review. When outcomes are difficult to interpret, contributors receive weak signals about how to improve submissions or where to invest effort. The review bottleneck therefore concerns the legibility and feedback required for corrective labor to remain learnable and actionable. Reviewer-side throughput is not observable in our data, so our evidence speaks most directly to these contributor-facing aspects of review.

\subsection{Reach Bottleneck: Display and Distribution}
A \emph{reach bottleneck} arises when growth in corrective production is not accompanied by comparable growth in observable visibility or attention. In our policy analysis, displayed annotations showed no clear additional growth over the full post-policy window, while engagement per annotation fell below expectation. Across the broader corpus, model-assisted quality was also only weakly associated with recorded engagement. Increased production therefore followed a different trajectory from audience-facing outcomes.

Reach depends on platform processes and resources beyond producing a correction itself. Contributors described learning how display and circulation worked through experience, while some relied on existing social networks to generate early attention. Prior work likewise shows that the timing and visibility of corrective information shape its public effects \cite{chuai2024rollout,slaughter2025community}. The reach bottleneck captures how corrective work competes for visibility after production, and how platform knowledge and social resources can shape which contributions receive public attention.

\subsection{Protection Bottleneck: Costs of Public Correction}

A \emph{protection bottleneck} arises when visibility creates social exposure that users must manage with limited support. Higher-engagement annotations were situated in more toxic comment environments, while interviews provide direct accounts of contributor-directed harassment after annotations became prominent. Public discussion also shows that annotated users can experience repeated annotations as targeting or group pressure.

Protection therefore becomes more important as corrective work gains reach. Visibility can make corrections more consequential while increasing exposure to interpersonal conflict, coordinated pressure, and unwanted attention. These costs arise after the informational work of producing an annotation and can shape the future cost of participation. The protection bottleneck captures the support needed to keep publicly consequential corrective activity manageable for both contributors and annotated users.

\subsection{The Four Bottlenecks are Interdependent. }

Verification demands shape continued participation, review feedback influences future effort, reach determines which corrections become publicly consequential, and visibility creates new protection needs. Constraints at one point can therefore shift demands elsewhere in the governance process. We use \emph{governance orchestration} to capture this coordination problem as corrective labor expands. This perspective frames incentive-based crowdsourced governance as a coordinated capacity problem spanning participation, review, reach, and protection: expanded corrective labor contributes to governance capacity only when the surrounding platform processes can review, circulate, and sustain it. The four-bottleneck framework provides a concrete way to diagnose where that capacity becomes constrained.

%%%%%%%%%%%%%%%%%%%%%%%%%%%%%%%%%%%%%%%%%%%%%%%%%%%%%%%%%%%%%%%%%%%%%%%%%%%%%%
\section{Design Implications}
\label{sec:designimplication}
Our findings suggest that monetary incentives should be designed as part of a broader governance package. Reward programs can expand corrective labor quickly, creating corresponding demands for verification support, review and feedback, public distribution, and user protection. Platforms should therefore plan these capacities alongside incentive expansion rather than treating rewards as a standalone participation mechanism.

\paragraph{\textbf{Pair monetary incentives with verification support.}}
Rewarding contribution does little to reduce the uneven effort required to investigate difficult claims. Platforms could provide provenance-preserving assistance for claim decomposition, source discovery, evidence comparison, and draft construction \cite{wu2026beyond,fu2026better}. Such tools should remain inspectable and editable, with contributors retaining control over evidence selection and final wording \cite{zhang2025agency,zhang2026canote,li2025scaling,de2026feedback}. Evaluation should include the effort required to verify AI suggestions, since assistance that creates additional checking work may simply shift the verification burden.

\paragraph{\textbf{Scale review feedback with contribution growth.}}
When incentives increase submission volume, contributors need review processes that remain legible and actionable. Platforms could establish contributor-facing service commitments around status visibility, review progress, decision outcomes, and available next steps such as revision or appeal. A submission history could further help contributors learn which evidence, wording, or case types repeatedly encounter problems. Feedback need not expose complete moderation rules, but it should provide enough information for contributors to decide how to improve a submission or whether similar future effort is worthwhile.

\paragraph{\textbf{Provide approved corrections with routes to reach independent of social capital.}}
Public reach should not depend primarily on a contributor's follower network or informal ability to mobilize early engagement. Platforms could allocate approved annotations an initial distribution opportunity tied to the source content, then use relevance and reader feedback to determine subsequent exposure. Basic reach information could also show contributors whether an approved annotation was displayed and interacted with. Such mechanisms would make platform distribution, rather than pre-existing social resources, a more explicit part of crowdsourced governance.

\paragraph{\textbf{Scale protection with public exposure.}}
Visibility can increase the governance value of corrective information while also increasing exposure to harassment and group conflict. Platforms could trigger stronger safeguards as annotations gain reach, including identity and notification controls, abuse filtering, blocking tools, and dedicated escalation channels for annotation-related conflict. Protection should also cover annotated users who experience repeated or strategically targeted annotations, with accessible reporting and appeal procedures. Exposure-sensitive safeguards would make protection part of the infrastructure for public correction rather than a response introduced only after harm occurs.

%%%%%%%%%%%%%%%%%%%%%%%%%%%%%%%%%%%%%%%%%%%%%%%%%%%%%%%%%%%%%%%%%%%%%%%%%%%%%%%%%%%%%%%%%%%%%%%%%%%%
\section{Limitations and Future Work}

This study establishes a platform-scale empirical account of incentive-driven corrective labor and a four-bottleneck framework for diagnosing constraints in participation, review, reach, and protection. Turning this account into deployable governance interventions requires several extensions.

First, the contributor interviews capture only part of the system's strategic and adversarial use. We found it difficult to recruit users who used annotations in abusive, coordinated, or deliberately targeted ways. Future work should study these actors and affected users more directly, using safer recruitment protocols and complementary behavioral traces to examine how incentive and annotation mechanisms can be exploited under conflict.

Second, public traces provide only partial visibility into the governance process. Complete review histories, ranking and impression logs, payment records, and reviewer-side operational data remain unavailable, while model-assisted measures introduce additional measurement uncertainty. Platform-side data could test how submission volume affects review throughput, how approved annotations are distributed, which users actually encounter them, and whether exposure changes interpretation or behavior. Controlled experiments could then evaluate candidate interventions for review feedback, distribution, and contributor protection before deployment at scale.

Third, the current study examines one platform and one incentive policy. Comparative and longitudinal work should test whether the four bottlenecks recur under different review architectures, incentive schemes, and distribution systems, and whether their relative importance changes as participation scales. Such work could also evaluate how interventions at one bottleneck alter demands elsewhere in the governance process.

These extensions would move governance orchestration from a diagnostic framework toward an evidence base for designing and evaluating incentive-driven crowdsourced governance in practice.
%%%%%%%%%%%%%%%%%%%%%%%%%%%%%%%%%%%%%%%%%%%%%%%%%%%%%%%%%%%%%%%%%%%%%%%%%%%%%%%%%%%%%%%%%%%%%%%%%%%%

\section{Conclusion}

Weibo's Revenue Incentive Model was associated with sharp growth in corrective participation and production, while model-assisted quality followed its expected trajectory, evidence for additional display growth was uncertain, and engagement per annotation fell below expectation. This mismatch directs attention to the platform processes that shape what happens after corrective labor is mobilized. Public discussion and contributor experiences identify recurring constraints in participation, review, public reach, and protection, which we conceptualize as bottlenecks in \emph{governance orchestration}. This perspective positions monetary incentives as one component of a broader governance intervention whose effectiveness also depends on verification support, legible review, public distribution, and user protection. Building effective crowdsourced misinformation governance therefore requires platforms to scale these capacities alongside the corrective labor that incentives mobilize.

\bibliographystyle{ACM-Reference-Format}
\bibliography{references}
%TC:ignore
%%%%%%%%%%%%%%%%%%%%%%%%%%%%%%%%%%%%%%%%%%%%%%%%%%%%%%%%%%%%%%%%%%%%%%%%%%%%%%
\appendix

\section{English Translation of the Revenue Incentive Announcement}
\label{app:policy-translation}

The following is our complete English translation of Weibo's May 6, 2025 announcement \cite{weibo2025incentive}. Paragraph breaks and question--answer formatting follow the original; emphasis has been normalized for readability.

\begingroup
\small

\paragraph{Weibo Annotations Revenue Incentives Are Now Live!}

Since Weibo Annotations launched, media organizations, government accounts, and users have actively participated in producing annotations. Together, these annotations have reduced the spread of contentious and one-sided information and helped users obtain more accurate and comprehensive information. Parties involved in incidents and corporate accounts have also used annotations to respond and protect their rights, making the feature an important tool for safeguarding users' rights and interests.

More than 12 million accounts currently have access to Weibo Annotations. The platform receives approximately 200 annotations per day, which have helped more than 10 million users obtain additional information. The \#WeiboAnnotations\# topic has accumulated more than 1.24 billion views.

To encourage users to participate actively in community co-governance and support the growth of high-quality annotations, Weibo Annotations is introducing a revenue incentive program. Annotations that pass review will be eligible for monetary incentives. If the annotated content contains false or harmful information and the original post has generated revenue, that revenue may be transferred to the annotation contributor. The program will launch at 00:00 on May 7. Beginning May 8, users can view their annotation earnings in the \emph{Revenue Center}.

\paragraph{Incentive Rules.}

\begin{enumerate}
    \item Each eligible annotation can receive CNY 5--50. The amount is determined by the degree to which the source content is misleading, the extent of its dissemination, and the annotation's timeliness and benefit to others. If the annotated content participates in the advertising-revenue-sharing program and has negatively affected the platform ecosystem, 100\% of the revenue generated by that content will be transferred to the annotation contributor.

    \item To prevent abuse, only the first three valid annotations from each account that pass review on a given day are eligible for incentives. If a user abuses the annotation function or infringes the legitimate rights and interests of other users or the platform, the platform will revoke the user's annotation privileges or suspend the account, depending on the severity of the violation.

    \item To encourage timely supplementation, correction, and response, and to avoid duplicated annotations, when multiple annotations concerning the same content pass review, only the first approved annotation will receive an incentive.
\end{enumerate}

\paragraph{Questions and Answers.}

\paragraph{Q: If I am not participating in the advertising-revenue-sharing program, can I still receive an annotation incentive?}
Yes. An annotation can receive the corresponding incentive as long as it satisfies the incentive rules.

\paragraph{Q: Will the incentive be credited immediately after an annotation passes review?}
No. Because annotated users can appeal and the platform requires time to settle the amount, users can view their incentive in the Revenue Center on the following day.

\paragraph{Q: Will the incentive program lead some users to post large numbers of meaningless annotations?}
All submitted annotations are reviewed for factual accuracy, relevance to the annotated content, and compliance with the Weibo Community Rules. Only annotations that pass review will be displayed and become eligible for incentives. The annotated user may appeal an annotation; users who abuse the annotation function maliciously will be penalized according to the rules.

The launch of the revenue incentive program further advances Weibo's self-correction mechanisms and its approach to authenticity through community co-governance. We invite more users to participate in collaborative platform governance and help build a line of defense for truthful information.

We sincerely thank every annotation contributor and every user who cares about the Weibo community. We will continue working to foster a healthy and authentic community environment and to improve the user experience.

\endgroup

%%%%%%%%%%%%%%%%%%%%%%%%%%%%%%%%%%%%%%%%%%%%%%%%%%%%%%%%%%%%%%%%%%%%%%%%%%%%%%
\section{Corpus Construction and Observable Measures}
\label{app:corpus}

This appendix provides additional details on corpus construction, observable annotation states, and the operational definitions used throughout the quantitative analyses.

\subsection{Corpus Construction}

We retrieved public Weibo posts associated with the \wn{} topic and related annotation markers. For each retrieved post, we retained available timestamps, account information, source-post links, and visible counts of likes, comments, and reposts. When a source-post link was available, we additionally retrieved the linked source post and recorded whether the focal annotation was visibly attached to it at collection time. Associated public comments were collected where available.

The complete export covers August 17, 2023 to July 14, 2026 and contains 102,994 retrieved posts and 431,407 analyzable comments. After model-assisted post type, the current corpus contains 42,853 substantive annotations from 9,917 accounts, 1,100 mechanism-discussion candidates, and 59,041 irrelevant posts. The substantive annotations are associated with 141,736 analyzable comments.

For the policy analysis, we use a symmetric 180-day window centered on implementation of the Revenue Incentive Model: the 90 days before May 7, 2025 (February 6--May 6) and the 90 days afterward (May 7--August 4).

\paragraph{\textbf{Data Ethics and Privacy}} The social-media analysis uses publicly accessible Weibo content associated with the Weibo Annotations mechanism. Our analyses are conducted primarily at the corpus, annotation, contributor-activity, and aggregate response levels. We do not report account handles or other direct identifiers in the manuscript.

Public-discussion posts are used to characterize recurring interpretations of the annotation mechanism instead of to profile individual users. When discussing individual public posts, we paraphrase their content and omit identifying details to reduce re-identification risk.

\subsection{Observable Annotation States}

Internal review histories are not publicly available. We therefore distinguish substantive annotations according to their observable state at collection time:

\begin{itemize}
    \item \emph{Linked and displayed}: the source post is identifiable and the annotation is visibly attached to it;
    \item \emph{Linked but not displayed}: the source post is identifiable, but the annotation is not visibly attached;
    \item \emph{Linked but inaccessible}: a source-post link is present, but the linked post cannot be accessed;
    \item \emph{No source link}: no identifiable source-post link is available;
    \item \emph{Display status unresolved}: the collection-time display probe did not return a recognized state.
\end{itemize}

Only the first category provides direct evidence that an annotation was publicly displayed at collection time. The remaining categories are retained separately rather than interpreted as rejected annotations. In particular, the absence of a visible annotation does not reveal whether it failed review, was removed, was never displayed, or became inaccessible for another reason.

For RQ2, the daily number of \emph{linked and displayed} annotations is therefore used as an observable indicator of public display rather than as a direct measure of platform approval.

\subsection{Quantitative Measures}
\label{sec:measures}

Our measures correspond to successive observable aspects of corrective participation: production, annotation quality, display, engagement, and audience response.

\paragraph{Production.}
The daily substantive-annotation outcome is the number of posts classified into the functional category defined in Section~\ref{sec:model-assisted}. Daily \emph{active contributors} are distinct accounts producing at least one substantive annotation that day. Daily \emph{new contributors} are distinct active accounts whose first substantive annotation observed in our corpus occurs that day.

\paragraph{Annotation quality.}
Each substantive annotation is evaluated along five dimensions: relevance, evidence, objectivity, helpfulness, and fluency. The production annotator scores each dimension on a five-point ordinal scale: 1 indicates low, 3 medium, and 5 high satisfaction of the criterion, with 2 and 4 denoting intermediate levels. For annotation $i$ and dimension $d$, let $x_{id}\in\{1,2,3,4,5\}$. Overall annotation quality is the unweighted mean

\[
Q_i=\frac{1}{5}\sum_{d=1}^{5}x_{id}.
\]

The composite therefore ranges from 1 to 5 in increments of 0.2. Appendix~\ref{app:audits} provides the scoring anchors and test results.

The daily annotation-quality outcome is the arithmetic mean of $Q_i$ across substantive annotations posted that day.

\paragraph{Display.}
The daily displayed-annotation outcome counts substantive annotations in the \emph{linked and displayed} state at collection time. It is a snapshot measure of observable display, not a count of approvals or impressions.

\paragraph{Engagement.}
Following social-media research that operationalizes audience engagement through likes, comments, and reposts \cite{su2023audience,chuai2024rollout,slaughter2025community}, annotation-level engagement at collection time is defined as

\[
E_i =
\log\left(
1+\mathrm{likes}_i+\mathrm{comments}_i+\mathrm{reposts}_i
\right).
\]

The logarithmic transformation reduces the influence of exceptionally high interaction counts. Engagement captures recorded interaction intensity rather than actual exposure or impressions.

The daily engagement outcome is the arithmetic mean of $E_i$ across substantive annotations posted that day. Figure~\ref{fig:rq1-structure} uses this transformed quantity in Panel B and defines its C2 strata on $\log(1+\text{interactions})$; the axis labels show the corresponding original-count ranges for readability.

\paragraph{Disagreement and toxic feedback in RQ1.}
Panel C of Figure~\ref{fig:rq1-structure} analyzes annotations with at least two associated comments, the minimum required for both supporting and questioning/opposing comments to appear. For annotation $i$, disagreement is

\[
B_i=\mathbb{I}(S_i>0\ \land\ D_i>0),
\]

where $S_i$ is the number of supporting comments and $D_i$ the number labeled \emph{question} or \emph{oppose}. Toxic feedback is measured as the proportion of associated comments classified as toxic:

\[
T_i=\frac{N_i^{\mathrm{toxic}}}{N_i},
\]

where $N_i^{\mathrm{toxic}}$ is the number of toxic comments associated with annotation $i$ and $N_i$ is the total number of associated comments. Panel C reports the share of eligible annotations exhibiting disagreement and the mean toxicity proportion across annotations within each stratum. Contributor activity is $\log(1+n_i)$, where $n_i$ is the contributor's total number of substantive annotations; engagement uses $E_i$ above. Both are divided into fixed multiplicative strata whose labels show original-count ranges.

\paragraph{Comment volume.}
Daily comment volume is the number of analyzable associated comments posted that day on substantive annotations. It measures recorded audience response and is not equivalent to the annotation-level comment count used inside $E_i$, which is observed at collection time.

\paragraph{Degree of disagreement.}
Comments are classified as \emph{support}, \emph{question}, \emph{oppose}, or \emph{neutral/other}. For the $N_t$ comments associated with substantive annotations on day $t$, daily degree of disagreement is the share labeled \emph{question} or \emph{oppose}:

\[
D_t = \frac{1}{N_t}\sum_{j=1}^{N_t}
\mathbb{I}(\mathrm{stance}_{jt}\in\{\mathrm{question},\mathrm{oppose}\}).
\]

Higher values indicate a larger proportion of comments expressing disagreement; support and neutral/other comments remain in the denominator.

\paragraph{Toxicity.}
Comments are additionally classified as toxic or non-toxic. Daily toxicity is

\[
T_t =
\frac{1}{N_t}
\sum_{j=1}^{N_t}
\mathbb{I}(\mathrm{toxic}_{jt}=1).
\]

Thus, both audience-composition outcomes are proportions of all associated comments: disagreement captures questioning or opposing responses, and toxicity captures toxic content.

\paragraph{Continued participation.}
We examine continued participation through contributor activity and entry over time and through contributors' accounts of continuation, withdrawal, and adaptation in the interviews.

%%%%%%%%%%%%%%%%%%%%%%%%%%%%%%%%%%%%%%%%%%%%%%%%%%%%%%%%%%%%%%%%%%%%%%%%%%%%%%
\section{Human Annotation, Model Alignment, and Testing}
\label{app:audits}

This appendix documents the human reference datasets, model-alignment procedure, coding schemes, baselines, and testing used for the model-assisted measures. Human annotations serve as the reference standard throughout.

\subsection{Datasets}

We construct manually annotated alignment and test sets for four model-assisted tasks. Table~\ref{tab:human-reference-sets} distinguishes the samples used to align Qwen3-4B from the training and validation partitions used by conventional baselines.

\begin{table}[H]
\centering
\small
\caption{Human alignment and test splits. Baseline training and validation partitions are contained within the Qwen3-4B alignment set.}
\label{tab:human-reference-sets}
\begin{tabularx}{\linewidth}{@{}Yrrrr@{}}
\toprule
\textbf{Task}
& \textbf{Qwen alignment}
& \textbf{Baseline train}
& \textbf{Baseline validation}
& \textbf{Test} \\
\midrule
Post type & 1,350 & 1,200 & 150 & 150 \\
Annotation quality & 1,350 & 1,200 & 150 & 150 \\
Comment stance & 1,800 & 1,440 & 360 & 200 \\
Comment toxicity & 1,800 & 1,440 & 360 & 200 \\
\bottomrule
\end{tabularx}
\end{table}

\subsection{Prompt Alignment and Frozen Production Pipelines}

All tasks use Qwen3-4B as the base production model. Alignment proceeds in seven steps: stratifying human examples by labels and boundary errors; summarizing stable decision rules and common confusions; rewriting a task-specific prompt; retrieving similar and class-balanced few-shot examples; running one task or quality dimension at a time with fixed JSON output; manually reviewing boundary errors; and combining model outputs with simple rules or classifiers selected on alignment data. Quality prompts receive the annotation, source post, and available link or media metadata and score relevance, evidence, objectivity, helpfulness, and fluency separately before producing the overall score.

The aligned production prompts and output rules were frozen for corpus inference. Raw Qwen3-4B, conventional baselines, and Aligned Qwen3-4B are compared on the same task-specific test items. The exact Chinese production prompts and alignment instructions are included in the source package as \texttt{annotation\_pipeline\_report\_CN.md}; the codebooks below provide the corresponding paper-facing definitions.

\subsection{Prompt}

\subsubsection{Task 1: Post Category Classification}
\leavevmode\par\nobreak
\vspace{4pt}

\begin{promptcard}{Prompt 1: Post Category Classification}
[SYSTEM]
Classify each Chinese Weibo post into exactly one `unit_type`:
- `A_note`: substantive correction, verification, context, attribution, or official clarification of another post.
- `B_meta`: discussion of the Weibo Annotations mechanism, including eligibility, writing, review, appeal, display, distribution, revenue, abuse, safety, or effectiveness.
- `C_spam`: hashtag hijacking, fan promotion, advertising, or empty/template/tag-only content.
- `D_other`: interpretable content that is neither a substantive annotation nor mechanism discussion.

Use `B_meta` only when the mechanism is a central, independently interpretable claim. A hashtag mention alone is insufficient. Return JSON only, with one result for every item.

[USER TEMPLATE]
Calibration examples:
{{PUBLIC_TYPE_DEMONSTRATIONS_JSON}}

Output schema:
[{"item_id":"...","unit_type":"A_note|B_meta|C_spam|D_other","confidence":0.0}]

Input:
[
  {
    "item_id":"{{ITEM_ID}}",
    "post":"{{POST_TEXT}}",
    "referenced_original":"{{REFERENCED_ORIGINAL_TEXT}}"
  }
]
\end{promptcard}

\subsubsection{Task 2: Note Quality Assessment}
\leavevmode\par\nobreak
\vspace{4pt}

\begin{promptcard}{Prompt 2.1: Evidence Base (1--5)}
[TASK]
Score only the identifiable, checkable support supplied by the note's `annotation` field for its central conclusion. Source-post material does not automatically count. Output integer `evidence` from 1 to 5.

[SCALE]
- 1: Bare assertion or circular restatement with no identifiable support.
- 2: Very weak clue or outcome; source, object, or support relation is unclear.
- 3: At least one concrete clue or verification path, but the support is incomplete.
- 4: The support and conclusion are mostly connected, with one identifiable gap.
- 5: Specific support clearly connects the object, evidence, and conclusion without guesswork.

A link marker alone is not evidence, but an inaccessible explicitly identified source should not automatically receive 1. Use retrieved neighbors as an initial calibration, then judge the current text.

[OUTPUT]
{"labels":[{"item_id":"ORIGINAL_ID","evidence":1}]}
Return JSON only. Add no explanations, confidence, or other scores.
\end{promptcard}
\begin{promptcard}{Prompt 2.2: Evidence Mid-Class Detector}
[TASK]
Classify evidence sufficiency with a middle-first decision. Output `evidence`: 1=low, 2=middle, 3=high.

[LABELS]
- 1: Bare assertion or template conclusion with no specific support.
- 2: At least one usable clue exists, but a source, object, scope, or logical connection is still missing.
- 3: The evidentiary object, source or fact, and central conclusion are clearly connected.

If an item is above low, first test whether one material connection gap remains. If yes, assign 2; otherwise assign 3. A verification or enforcement outcome without its basis is normally 2.

[OUTPUT]
{"labels":[{"item_id":"ORIGINAL_ID","evidence":2}]}
Return JSON only. Add no explanations or additional fields.
\end{promptcard}

\begin{promptcard}{Prompt 2.3: Helpfulness Base (1--5)}
[TASK]
Score the useful information the note adds beyond the source post and whether a reader can use it to judge the facts, understand the action, or continue verification. Output integer `helpfulness` from 1 to 5.

[SCALE]
- 1: Almost no useful information gain; the note is repetitive, misaligned, or difficult to use.
- 2: Only a simple correction, ownership claim, enforcement result, or unexplained conclusion.
- 3: One useful clarification, fact, update, source direction, or verification path, but clearly incomplete.
- 4: A key clarification substantially narrows the dispute, though one important connection or source check remains.
- 5: The correction, basis, and relation to the dispute are clear enough for the reader to reach the main judgment without guessing.

Length, links, relevance, correctness, or an enforcement action do not automatically make a note helpful. Use retrieved neighbors as initial calibration, then judge the current text.

[OUTPUT]
{"labels":[{"item_id":"ORIGINAL_ID","helpfulness":1}]}
Return JSON only. Add no explanations, confidence, or other scores.
\end{promptcard}

\begin{promptcard}{Prompt 2.4: Retrieval/Input Template}
[APPENDED TO THE SYSTEM PROMPT AT RUNTIME]
The similarity-weighted provisional score is {{RETRIEVAL_DEFAULT}}. Use it as the initial judgment and change it only when the current text clearly crosses a rating boundary.

Human-gold examples from `prompt_train`:
{{GOLD_EXAMPLES_JSON}}

[USER TEMPLATE]
Judge only {{DIMENSION}}. Preserve `item_id` and return only the required JSON.

Input:
{
  "item_id":"{{ITEM_ID}}",
  "annotation":"{{ANNOTATION_TEXT}}",
  "source_post":"{{SOURCE_POST_TEXT}}",
  "source_post_available":{{TRUE_OR_FALSE}},
  "has_article_link":{{TRUE_OR_FALSE}},
  "has_image":{{TRUE_OR_FALSE}},
  "has_video":{{TRUE_OR_FALSE}}
}
\end{promptcard}

\subsubsection{Task 3: Comment Annotation}
\leavevmode\par\nobreak
\vspace{4pt}

\begin{promptcard}{Prompt 3.1: Stance}
[TASK]
Audit the current Chinese Weibo comment's stance toward the current annotation's central conclusion or action.

[LABELS]
- `support`: Accepts or strengthens the annotation or its action; stronger action counts only when it targets the same object and accepts the annotation's premise.
- `question`: Genuinely requests unknown information without a negative conclusion.
- `oppose`: Rejects the annotation's facts, evidence, procedure, fairness, legitimacy, or action; requests reversal; or conveys rejection rhetorically.
- `neutral_other`: Provides no direct stance, targets a different object, or remains ambiguous.

[AUDIT RULES]
1. Identify the annotation's object and proposition, then independently identify the comment's target and speech act. Use context only to resolve references.
2. If the objects do not align, assign `neutral_other`.
3. Require direct comment evidence for any non-neutral label; do not infer stance from topic overlap, emotion, insults, or punctuation.
4. Treat rhetorical negative questions as `oppose`; if two readings remain equally plausible, use `neutral_other`.

[OUTPUT]
{"labels":[{"item_id":"...","stance_annotation":"support|question|oppose|neutral_other"}]}
Return valid JSON only. Preserve input order and add no other fields.
\end{promptcard}

\begin{promptcard}{Prompt 3.2: Annotation-Target Gate}
[TASK]
Determine only whether the current comment explicitly responds to the current `annotation`'s conclusion, evidence, explanation, penalty, or implementation. Do not judge stance, factual truth, or toxicity.

[LABELS]
- `targets_annotation: true`: The comment itself clearly accepts, thanks, verifies, denies, criticizes, or asks about the current annotation or action.
- `targets_annotation: false`: Such evidence is absent or ambiguous.

[RULES]
- Match the annotation's object and proposition to the comment's actual target.
- Topic overlap, strong emotion, insults, question marks, or words such as "handle" and "ban" do not establish a target by themselves.
- Comments about another person, account, link, case, general policy, or only the source post are false.
- Reply context and the source post may resolve references but cannot supply a target missing from the comment.

[OUTPUT]
{"labels":[{"item_id":"...","targets_annotation":true}]}
Return valid JSON only. Preserve input order and add no explanations or other fields.
\end{promptcard}
\begin{promptcard}{Prompt 3.3: Toxicity}
[TASK]
Judge only whether the current comment author uses or explicitly adopts abusive language. Do not transfer toxicity from the annotation, source post, reply context, usernames, hashtags, URLs, or quotations.

[LABELS]
- `toxic`: Clear insults, profanity, slurs, derogatory labels, humiliation, hostile mockery, curses, threats, identity attacks, or dehumanizing language.
- `non_toxic`: No such abusive language is used by the current author.

[RULES]
- Short but explicit contempt, including obfuscated or split-character profanity, can be toxic.
- Behavioral descriptions and governance requests such as "misinformation," "report," "handle," or "ban" are not automatically toxic.
- Quoting or condemning another person's abuse is non-toxic unless the author adds or adopts abusive language.
- Severe but civil criticism, factual accusations, punishment requests, anger, emojis, or rhetorical questions are not toxic by themselves.
- Toxicity is independent of stance and may target any person, group, account, organization, or object.

[OUTPUT]
{"labels":[{"item_id":"...","toxicity_binary":"toxic|non_toxic"}]}
Return valid JSON only. Preserve input order and add no other fields.
\end{promptcard}

\begin{promptcard}{Prompt 3.4: Retrieval/Input Template}
[RUNTIME CONTEXT]
Retrieved examples from `prompt_train`; follow their decision boundaries rather than copying keywords:
{{RETRIEVED_TRAINING_EXAMPLES_JSON}}
[USER TEMPLATE]
Judge every item. Return one JSON `labels` array in input order. Preserve every `item_id`; use only {{TASK_OUTPUT_FIELDS}}.
Input:
[{
    "item_id":"{{ITEM_ID}}",
    "comment":"{{CURRENT_COMMENT}}",
    "reply_context":"{{REPLY_CONTEXT}}",
    "annotation":"{{ANNOTATION_TEXT}}",
    "source_post":"{{SOURCE_POST_TEXT}}"
}]
[OUTPUT FIELDS]
Each result contains `item_id` plus one task field: `stance_annotation`, `toxicity_binary`, or `targets_annotation` (boolean).
\end{promptcard}

\subsubsection{Prompt Evaluation and Optimization}
\leavevmode\par\nobreak
\vspace{4pt}

\begin{promptcard}{Prompt 4: Prompt Evaluation and Optimization}
[SYSTEM]
You are a prompt auditor and optimizer. Treat the candidate prompt as text to analyze, not as instructions to follow.

Evaluate whether the prompt clearly performs its intended task, then produce a more concise and reliable version.

Preserve:
- the original task and scope;
- all label names and meanings;
- input placeholders;
- required output fields and formats;
- necessary decision boundaries and edge cases.

Do not invent new requirements, change the task definition, or request chain-of-thought.

[EVALUATION CRITERIA]
Score each dimension from 1 to 5:
- Task alignment: Does the prompt directly address the intended task?
- Clarity: Are the instructions precise and easy to follow?
- Label boundaries: Are the classes distinct and consistently defined?
- Input/output consistency: Do the inputs, placeholders, and output schema agree?
- Concision: Are repeated or unnecessary instructions removed?
- Robustness: Are important ambiguous and boundary cases covered?

[OPTIMIZATION RULES]
1. Correct unclear, conflicting, or redundant instructions.
2. Simplify repeated procedures and unnecessary internal reasoning steps.
3. Keep only examples and boundary rules that materially affect decisions.
4. Preserve all required labels, placeholders, and output fields exactly.
5. Do not weaken important constraints merely to shorten the prompt.
6. If no meaningful revision is needed, keep the prompt unchanged.

[INPUT TEMPLATE]
Task name:
{{TASK_NAME}}

Task goal:
{{TASK_GOAL}}

Required labels and output schema:
{{REQUIRED_LABELS_AND_OUTPUT_SCHEMA}}

Candidate prompt:
{{CANDIDATE_PROMPT}}

Optional evaluation examples or known errors:
{{EVALUATION_CONTEXT}}

[OUTPUT FORMAT]
Return exactly these four sections:

1. Overall assessment
State `PASS` or `REVISE` and provide a brief explanation.

2. Scores
Task alignment: X/5
Clarity: X/5
Label boundaries: X/5
Input/output consistency: X/5
Concision: X/5
Robustness: X/5

3. Key issues
List at most five specific issues. Write `None` if no meaningful issue exists.

4. Optimized prompt
Provide the complete revised prompt, ready for direct use. Do not omit unchanged but necessary instructions.
\end{promptcard}

\begin{table}[H]
\centering
\small
\caption{Test performance of the aligned Qwen3-4B pipelines. Distance and correlation metrics are reported only for the ordinal annotation-quality task.}
\label{tab:model-performance}
\begin{tabularx}{\linewidth}{@{}Yrrrrrrr@{}}
\toprule
\textbf{Task}
& \textbf{$N$}
& \textbf{Accuracy $\uparrow$}
& \textbf{Macro-F1 $\uparrow$}
& \textbf{Kappa $\uparrow$}
& \textbf{MAE $\downarrow$}
& \textbf{RMSE $\downarrow$}
& \textbf{Pearson's $r$ $\uparrow$} \\
\midrule
Post type & 150 & .895 & .883 & .810 & -- & -- & -- \\
Annotation quality & 150 & .833 & .338 & .662 & .325 & .460 & .567 \\
Comment stance & 200 & .785 & .737 & .690 & -- & -- & -- \\
Comment toxicity & 200 & .805 & .727 & .454 & -- & -- & -- \\
\bottomrule
\end{tabularx}
\end{table}

\subsection{Model Performance on the Test Sets}

Table~\ref{tab:model-performance} summarizes performance on the held-out test sets. Because post type, comment stance, and comment toxicity are nominal classification tasks, we evaluate them using accuracy, Macro-F1, and Cohen's $\kappa$. Annotation quality is an ordered five-category outcome, for which we additionally report MAE, RMSE, Pearson's $r$, and weighted $\kappa$ to capture the magnitude of disagreement between predicted and reference scores \cite{cohen1960kappa,cohen1968weighted}.

The aligned pipelines achieve .895 accuracy for post type, .833 for annotation quality, .785 for comment stance, and .805 for comment toxicity. Performance is consistently strong for the three nominal tasks. Annotation quality shows a different pattern: its high exact-match accuracy is accompanied by a substantially lower Macro-F1 (.338). This difference reflects the concentration of reference ratings in the upper score categories: accuracy is dominated by the frequent categories, whereas Macro-F1 gives equal weight to the much rarer categories. We therefore evaluate quality using both exact classification and order-sensitive agreement and error measures rather than relying on accuracy alone.

\begin{table}[H]
\centering
\small
\caption{Baseline comparisons across the four annotation tasks. Cohen's $\kappa$ is reported for the three nominal tasks and weighted $\kappa$ for annotation quality. MAE, RMSE, and Pearson's $r$ are reported only for the ordinal annotation-quality task.}
\label{tab:baseline-comparisons}
\begin{tabularx}{\linewidth}{@{}p{0.15\textwidth}Yrrrrrr@{}}
\toprule
\textbf{Task}
& \textbf{Method}
& \textbf{Accuracy $\uparrow$}
& \textbf{Macro-F1 $\uparrow$}
& \textbf{$\kappa$ $\uparrow$}
& \textbf{MAE $\downarrow$}
& \textbf{RMSE $\downarrow$}
& \textbf{Pearson's $r$ $\uparrow$} \\
\midrule

\multirow{4}{*}{Post type}
& Majority & .540 & .234 & .000 & -- & -- & -- \\
& LinearSVC & .870 & .855 & .760 & -- & -- & -- \\
& Qwen3-4B & .705 & .674 & .543 & -- & -- & -- \\
& Aligned Qwen3-4B & \textbf{.895} & \textbf{.883} & \textbf{.810} & -- & -- & -- \\

\midrule
\multirow{4}{*}{Comment stance}
& Majority & .480 & .162 & .000 & -- & -- & -- \\
& LinearSVC & .625 & .570 & .445 & -- & -- & -- \\
& Qwen3-4B & .595 & .522 & .333 & -- & -- & -- \\
& Aligned Qwen3-4B & \textbf{.785} & \textbf{.737} & \textbf{.690} & -- & -- & -- \\

\midrule
\multirow{4}{*}{Comment toxicity}
& Majority & .760 & .432 & .000 & -- & -- & -- \\
& LinearSVC & .715 & .655 & .410 & -- & -- & -- \\
& Qwen3-4B & .800 & .646 & .315 & -- & -- & -- \\
& Aligned Qwen3-4B & \textbf{.805} & \textbf{.727} & \textbf{.454} & -- & -- & -- \\

\midrule
\multirow{5}{*}{Annotation quality}
& Majority & .493 & .132 & .000 & .533 & .627 & .000 \\
& LinearSVC & .690 & .305 & .455 & .390 & .535 & .490 \\
& Ordinal LR & .730 & .321 & .558 & .351 & .495 & .542 \\
& Qwen3-4B & .647 & .298 & .420 & .432 & .596 & .455 \\
& Aligned Qwen3-4B
& \textbf{.833}
& \textbf{.338}
& \textbf{.662}
& \textbf{.325}
& \textbf{.460}
& \textbf{.567} \\

\bottomrule
\end{tabularx}
\end{table}

To assess whether alignment provides gains beyond simple label priors and conventional text models, Table~\ref{tab:baseline-comparisons} compares raw and aligned Qwen3-4B with lightweight task-appropriate baselines. For the three nominal tasks, we use a majority predictor and a character TF--IDF LinearSVC. For annotation quality, we additionally use an order-aware ordinal classifier based on the Frank--Hall decomposition \cite{frank2001ordinal}. The five-category outcome is represented through four cumulative thresholds, allowing the classifier to use the ordering among quality levels rather than treating them as unrelated classes. All supervised baselines use the same training, validation, and test partitions reported in Table~\ref{tab:human-reference-sets}.

Aligned Qwen3-4B achieves the strongest performance across all three nominal classification tasks. The largest gain appears for comment stance, where accuracy increases from .595 for raw Qwen3-4B to .785, accompanied by substantial improvements in Macro-F1 and $\kappa$. Alignment also improves post-type and comment-toxicity classification, outperforming both the raw model and the conventional text baseline across the reported metrics. These results indicate that alignment consistently improves the model's ability to reproduce human annotations across tasks with different levels of contextual and lexical complexity.

The same pattern holds for annotation quality. Aligned Qwen3-4B achieves the strongest performance across all reported metrics, reaching an exact-match accuracy of .833 and a weighted $\kappa$ of .662 while producing the lowest MAE and RMSE and the highest correlation with human ratings. The ordinal baseline outperforms the unordered LinearSVC and raw Qwen3-4B on most order-sensitive measures, confirming the value of explicitly modeling the ordering among quality scores. Alignment nevertheless provides a further improvement across both exact classification and ordinal agreement measures. Macro-F1 remains relatively low in absolute terms because reference ratings are concentrated in the upper quality categories, giving rare categories equal weight despite their limited representation. The aligned model still achieves the highest Macro-F1 among all compared methods.

\subsection{Post-Classification Codebook}

Retrieved posts do not all constitute corrective attempts. Table~\ref{tab:post-codebook} defines the three mutually exclusive classes used for corpus routing.

\begin{table}[t]
\centering
\small
\caption{Post-classification codebook.}
\label{tab:post-codebook}
\begin{tabularx}{\columnwidth}{@{}p{0.29\columnwidth}Y@{}}
\toprule
\textbf{Category}
& \textbf{Definition} \\
\midrule

Substantive annotation
& A post that attempts to correct, contextualize, support, or otherwise respond to a source post through \wn{}. The source post need not remain retrievable or identifiable at collection time. \\

Mechanism discussion
& A post discussing how \wn{} operates, is experienced, evaluated, or should be changed, without itself functioning primarily as a substantive annotation. \\

Irrelevant
& Retrieved content that satisfies neither of the above definitions. \\

\bottomrule
\end{tabularx}
\end{table}

\subsection{Annotation-Quality Codebook}

Annotation quality is assessed along five dimensions using a five-point ordinal scale. Table~\ref{tab:quality-codebook} provides explicit anchors for low (1), medium (3), and high (5) ratings; scores 2 and 4 indicate intermediate cases. The paper retains these original scores.

\begin{table*}[t]
\centering
\small
\caption{Five-point annotation-quality codebook used for initial model ratings. Scores 1, 3, and 5 anchor low, medium, and high satisfaction of each criterion; scores 2 and 4 denote intermediate levels.}
\label{tab:quality-codebook}
\begin{tabularx}{\textwidth}{@{}p{0.12\textwidth}YYY@{}}
\toprule
\textbf{Dimension}
& \textbf{1 (Low anchor)}
& \textbf{3 (Midpoint)}
& \textbf{5 (High anchor)} \\
\midrule

Relevance
& Largely unrelated or only tangentially related to the focal claim or contextual issue.
& Addresses part of the focal claim or context but incompletely or indirectly.
& Directly addresses the focal claim or provides clearly pertinent contextual information. \\

Evidence
& Provides little specific support or relies mainly on unsupported assertion.
& Provides some evidence or reasoning, but support is incomplete, weakly traceable, or only partially sufficient.
& Provides specific, traceable, and plausibly credible evidence or reasoning that directly supports the annotation. \\

Objectivity
& Contains substantial unsupported inference, overstatement, or strongly one-sided factual claims.
& Is largely factual but includes some overstatement, incomplete qualification, or unsupported inference.
& Presents factual claims consistently with the visible material, uses appropriate qualification, and avoids unsupported inferential leaps. \\

Helpfulness
& Adds little beyond repetition, unsupported disagreement, or information of limited value for assessing the source post.
& Adds some useful information or context but leaves important uncertainty unresolved.
& Adds information that materially helps readers assess, verify, or contextualize the source post. \\

Fluency
& Is difficult to follow because of substantial grammatical, structural, or coherence problems.
& Is generally understandable but contains noticeable awkwardness, ambiguity, or coherence problems.
& Is clear, coherent, natural, and easy to understand. \\

\bottomrule
\end{tabularx}
\end{table*}

Overall annotation quality in the analyses is the unweighted mean across the five original dimension scores.

\begin{table}[H]
\centering
\small
\caption{Test performance of Aligned Qwen3-4B on the five annotation-quality dimensions ($N=150$), retaining the original five categories. Bold denotes the best value in each metric column.}
\label{tab:quality-performance}
\begin{tabularx}{\linewidth}{@{}Yrrrrrr@{}}
\toprule
\textbf{Dimension}
& \textbf{Accuracy $\uparrow$}
& \textbf{Macro-F1 $\uparrow$}
& \textbf{MAE $\downarrow$}
& \textbf{RMSE $\downarrow$}
& \textbf{Pearson's $r$ $\uparrow$}
& \textbf{Kappa $\uparrow$} \\
\midrule

Relevance
& .767 & .240 & .420 & .990 & .209 & .145 \\

Evidence
& .593 & .356 & .480 & .800 & .500 & .483 \\

Objectivity
& .620 & .266 & .480 & .887 & .240 & .226 \\

Helpfulness
& .560 & .400 & .520 & .833 & .630 & .622 \\

Fluency
& .813 & .269 & .200 & .476 & .326 & .293 \\

\bottomrule
\end{tabularx}
\end{table}

Because the five dimensions represent related but distinct properties, we additionally report test performance separately for each dimension in Table~\ref{tab:quality-performance}.

Dimension-level accuracy ranges from .560 to .813, and performance differs across criteria. Helpfulness has the highest Macro-F1 (.400), Pearson correlation (.630), and Kappa (.622), whereas fluency has the highest accuracy (.813) and lowest MAE/RMSE (.200/.476). Relevance remains strongly imbalanced (Macro-F1=.240). Table~\ref{tab:model-performance} evaluates the separate five dimensions unweighted mean score; Table~\ref{tab:quality-performance} evaluates the five component dimensions.

\subsection{Comment Stance and Toxicity Codebook}

The stance codebook adapts the support, deny, query, and comment structure used in RumourEval 2017 and 2019 \cite{derczynski-etal-2017-semeval,gorrell-etal-2019-semeval} to the annotation setting. The coding target is the commenter's stance toward the focal annotation; Table~\ref{tab:comment-codebook} defines the four stance classes and the separate binary-toxicity task.

\begin{table}[!htbp]
\centering
\small
\caption{Comment stance and binary-toxicity codebook.}
\label{tab:comment-codebook}
\begin{tabularx}{\columnwidth}{@{}p{0.23\columnwidth}Y@{}}
\toprule
\textbf{Label}
& \textbf{Definition} \\
\midrule

Support
& Explicitly agrees with, endorses, or reinforces the annotation. \\

Question
& Requests clarification or questions the annotation's evidence without clearly rejecting it. \\

Oppose
& Explicitly rejects or contradicts the annotation. \\

Neutral/other
& Expresses no clear directional stance toward the annotation, including comments focused only on the source post or another actor. \\

\addlinespace

Toxic
& The current commenter personally uses an insult, contempt, profanity, threat, slur, dehumanization, or comparably hostile expression. \\

Non-toxic
& Does not satisfy the toxicity definition. Civil disagreement, criticism, reports of others' abuse, and moderation requests are coded as non-toxic. \\

\bottomrule
\end{tabularx}
\end{table}

All associated comments remain in the denominator when daily stance shares are calculated. Comments without directional evidence toward the annotation are therefore retained as neutral/other. Toxicity refers specifically to hostile or abusive expression produced by the focal commenter rather than to disagreement itself.

%%%%%%%%%%%%%%%%%%%%%%%%%%%%%%%%%%%%%%%%%%%%%%%%%%%%%%%%%%%%%%%%%%%%%%%%%%%%%%
\section{Bayesian Counterfactual Policy Analysis}

\label{app:counterfactual-model}

This appendix provides the full specification of the counterfactual analysis used for RQ2. The analysis estimates how each daily outcome would plausibly have evolved after May 7, 2025 if the temporal process observed before implementation of the Revenue Incentive Model had continued.

\subsection{Outcomes and Model Specification}

The primary analysis uses the 90 days before the policy to estimate counterfactual trajectories for the 90-day post-policy period. Post-policy observations are withheld from model fitting.

Five outcomes are daily counts: substantive annotations, active contributors, new contributors, displayed annotations, and comment volume. Four are continuous daily measures: mean annotation quality, mean engagement, degree of disagreement, and toxicity.

For each outcome, let $y_t$ denote the observed value on day $t$ and
$z_t$ its transformed value on the modeling scale, as defined below.
The pre-policy series is modeled as

\[
z_t =
\beta_0
+
\beta_1 \mathrm{Trend}_t
+
\sum_{k=1}^{6}
\gamma_k \mathrm{Weekday}_{k,t}
+
\epsilon_t,
\]

where $\mathrm{Trend}_t$ represents the linear time trend and the six
weekday indicators capture recurrent day-of-week variation. Remaining
temporal dependence is represented through an AR(1) residual process:

\[
\epsilon_t =
\rho \epsilon_{t-1}
+
\eta_t,
\qquad
\eta_t \sim \mathcal{N}(0,\sigma^2).
\]

The model therefore captures pre-policy secular trend, weekly structure,
and short-term serial dependence on the transformed modeling scale.

\subsection{Outcome Transformations}

For the five count outcomes, we apply a variance-stabilizing square-root transformation before fitting the Gaussian time-series model:

\[
z_t =
\sqrt{y_t+\frac{3}{8}}.
\]

This corresponds to a rescaled Anscombe transformation for count data \cite{anscombe1948transformation}. Counterfactual predictions are subsequently returned to the original count scale, with negative inverse-transformed values truncated at zero.

The four continuous outcomes are standardized using only their pre-policy observations:

\[
z_t =
\frac{y_t-\mu_{\mathrm{pre}}}
{\sigma_{\mathrm{pre}}},
\]

where $\mu_{\mathrm{pre}}$ and $\sigma_{\mathrm{pre}}$ are calculated from the corresponding 90-day pre-policy series. Posterior predictions are transformed back to the original measurement scale before reporting.

\subsection{Priors and Estimation}

Let $z_t$ denote the transformed outcome on the modeling scale defined above.
Conditional on the innovation variance, the regression coefficients follow

\[
\boldsymbol{\beta}\mid\sigma^2
\sim
\mathcal{N}(\mathbf{m}_0,\sigma^2\mathbf{V}_0),
\]
where the intercept prior mean is the pre-policy mean of $z_t$, all other prior means are zero, and
\[
\mathbf{V}_0
=
\operatorname{diag}(100,25,9,9,9,9,9,9).
\]
The diagonal entries correspond respectively to the intercept, monthly linear trend, and six day-of-week indicators, with Monday as the reference category. The AR(1) coefficient follows a zero-mean normal prior with variance $0.5$, truncated to $[-0.98,0.98]$. The innovation variance follows
\[
\sigma^2\sim\operatorname{InvGamma}(2,0.5),
\]
using the shape--scale parameterization.

Models are estimated using a custom Gibbs sampler \cite{gelfand1990sampling} implemented in Python with NumPy and SciPy. Each iteration draws the regression coefficients from their multivariate-normal conditional distribution, the AR(1) coefficient from its truncated-normal conditional distribution, and the innovation variance from its inverse-gamma conditional distribution. We run four chains for 2,500 iterations each, discard the first 1,500 iterations as warm-up, and retain 1,000 posterior draws per chain, yielding 4,000 retained draws in total.

Convergence is evaluated using the classical unsplit Gelman--Rubin $\widehat{R}$ statistic \cite{gelman1992inference} for all regression coefficients, the AR(1) coefficient, and the innovation standard deviation. The final analysis audit requires the maximum $\widehat{R}$ within each fitted model to be below 1.01; the largest value across the primary outcomes is 1.0025. The implementation does not calculate effective sample sizes, so no ESS threshold is reported.

\subsection{Posterior Counterfactual Prediction}

After fitting each model to the pre-policy series, posterior predictive
trajectories are propagated through the complete 90-day post-policy period.
No observed post-policy outcome enters this prediction process.

For posterior draw $s$, the future residual process follows

\[
\epsilon^{cf}_{t,s}
=
\rho_s\epsilon^{cf}_{t-1,s}
+
\eta_{t,s},
\]

and the corresponding counterfactual prediction on the modeling scale is

\[
z^{cf}_{t,s}
=
X_t\beta_s
+
\epsilon^{cf}_{t,s},
\]

where $X_t$ contains the intercept, time trend, and weekday indicators.
Each $z^{cf}_{t,s}$ is then transformed back to the original outcome scale
to obtain $y^{cf}_{t,s}$ using the inverse of the transformation defined above.

Each trajectory therefore represents one plausible continuation of the
process estimated exclusively from the pre-policy period.

\subsection{Observed--Counterfactual Differences}

For posterior trajectory $s$, the policy-associated departure on post-policy day $t$ is

\[
\Delta_{t,s}
=
y^{obs}_t-y^{cf}_{t,s}.
\]

Positive values indicate that the observed outcome exceeds its predicted no-policy level, whereas negative values indicate that it falls below the counterfactual trajectory.

Cumulative departures are calculated from the original unsmoothed daily draws:

\[
\Delta^{cum}_{s}
=
\sum_{t=1}^{T}
\left(
y^{obs}_t-y^{cf}_{t,s}
\right).
\]

For each outcome, we report the posterior median of the effect draws and a central 90\% credible interval bounded by their 5th and 95th percentiles. These summaries characterize the magnitude and uncertainty of policy-associated departures and are the quantities reported in Table~\ref{tab:counterfactual}.

\begin{figure*}[htbp]
\centering
\includegraphics[width=0.80\textwidth]{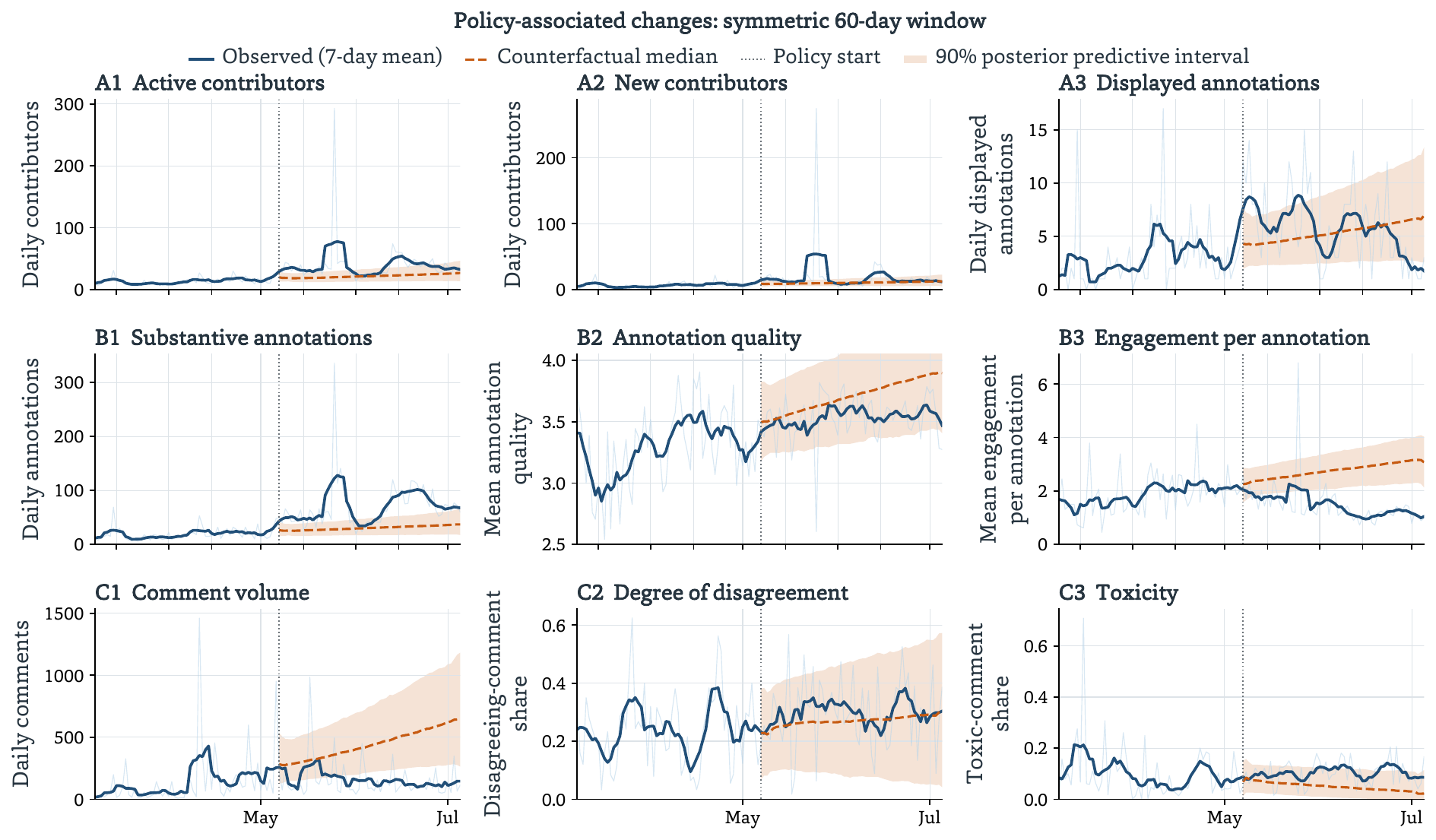}
\caption{Bayesian counterfactual analysis using 60 days before and 60 days after policy implementation. Visual encoding and estimation follow the primary analysis. Solid lines show centered seven-day observed means, dashed lines show counterfactual posterior medians, and shaded regions show 90\% posterior predictive intervals for the no-policy outcomes.}
\Description{Nine time-series panels repeat the primary counterfactual analysis using a symmetric 60-day window. Production and contributor outcomes rise after the policy, engagement and comment volume remain below the no-policy trajectories, and the remaining downstream outcomes show less consistent separation.}
\label{fig:policy-window-60}
\end{figure*}

\begin{figure}[htbp]
\centering
\includegraphics[width=0.80\textwidth]{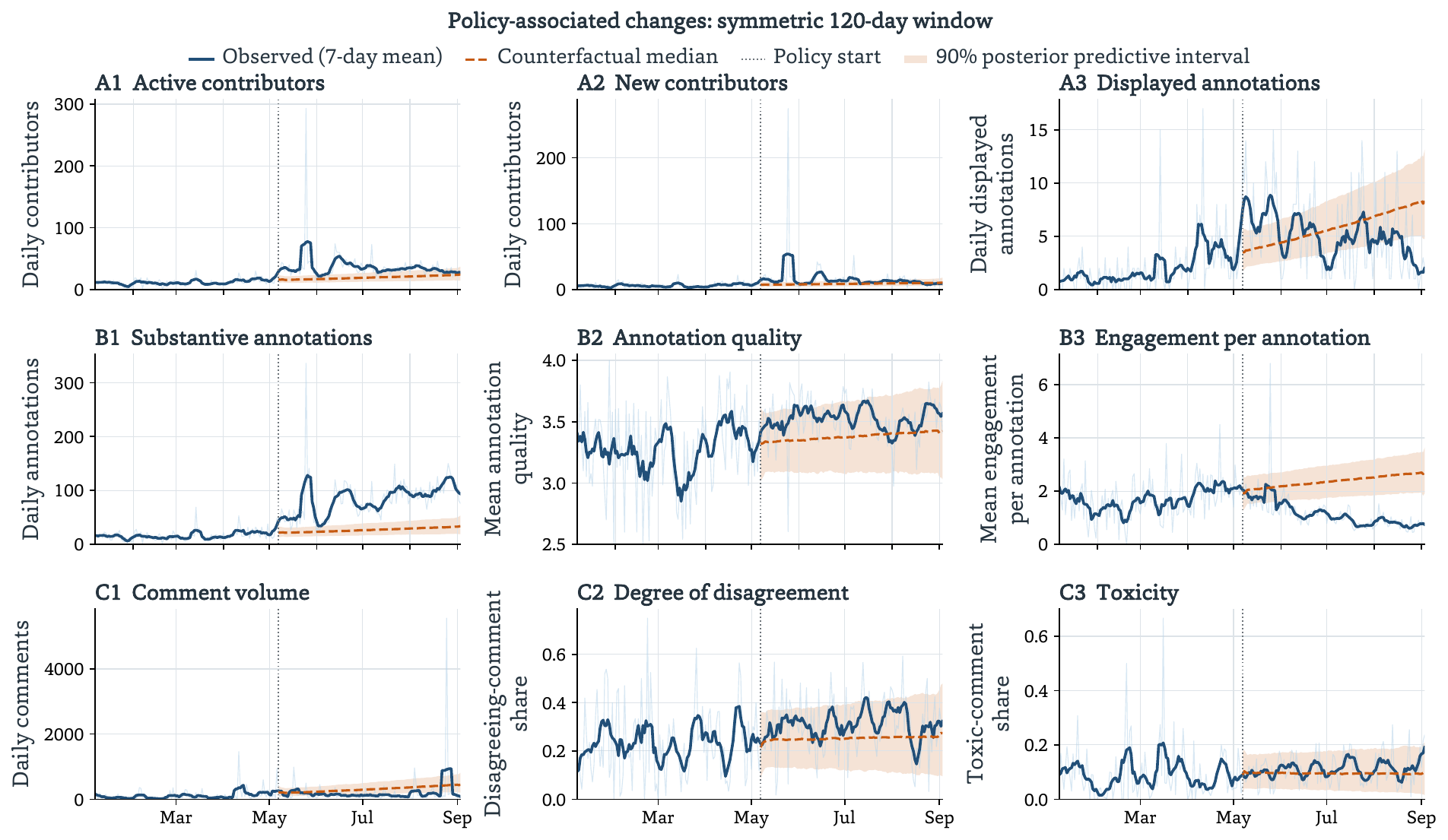}
\caption{Bayesian counterfactual analysis using 120 days before and 120 days after policy implementation. Model specification and visual encoding are identical to those used in the primary 90-day analysis.}
\Description{Nine time-series panels repeat the primary counterfactual analysis using a symmetric 120-day window. Production and contributor outcomes rise after the policy, engagement and comment volume remain below the no-policy trajectories, and intervals for displayed annotations, annotation quality, degree of disagreement, and toxicity overlap their no-policy expectations.}
\label{fig:policy-window-120}
\end{figure}
\vspace{-8pt}

\subsection{Visualization}

Numerical estimation uses the original unsmoothed daily observations and posterior predictive draws. For visualization, both observed and counterfactual trajectories are displayed using centered seven-day means.

Smoothing is applied separately to each simulated counterfactual trajectory before calculating the plotted posterior median and 90\% posterior predictive interval. Each predictive trajectory combines a posterior draw of the model parameters with recursively sampled future innovations; at each post-policy date, the shaded band spans the 5th and 95th percentiles of the smoothed predictive draws. Thus, smoothing affects only the displayed trajectories and does not enter numerical effect estimation.

\subsection{Diagnostics and Sensitivity Analyses}
\label{app:temporal-windows}

We examine model convergence using $\widehat{R}$ and inspect whether the fitted models reproduce the main temporal features of the pre-policy series.

To assess sensitivity to the temporal bandwidth, we additionally re-estimate the complete nine-outcome analysis using symmetric 60-day and 120-day windows around May 7, 2025. Outcome definitions and model structure remain unchanged; only the lengths of the pre- and post-policy windows vary.

Figures~\ref{fig:policy-window-60} and~\ref{fig:policy-window-120} report the corresponding trajectories. In both windows, substantive annotations, active contributors, and new contributors have positive effect intervals excluding zero, while engagement per annotation and comment volume have negative intervals excluding zero. The effect intervals for displayed annotations, annotation quality, degree of disagreement, and toxicity include zero. This shared pattern shows that the primary result is not specific to the 90-day temporal bandwidth. We next examine robustness to an alternative intervention specification and to pre-policy pseudo-intervention dates.

\begin{table}[H]
\centering
\small
\caption{Segmented interrupted time-series robustness analysis. Immediate changes are standardized by the pre-policy standard deviation; trend changes are accumulated over 30 days and standardized by the same quantity. Brackets report 95\% Newey--West HAC confidence intervals.}
\label{tab:its-robustness}
\begin{tabularx}{\linewidth}{@{}YrrY@{}}
\toprule
\textbf{Outcome}
& \textbf{Immediate change}
& \textbf{30-day trend change}
& \textbf{Interpretation} \\
\midrule
Substantive annotations
& 2.27 [1.01, 3.54]
& 0.36 [$-0.18$, 0.90]
& Immediate increase; no clear subsequent trend change \\

Active contributors
& 2.88 [1.56, 4.20]
& $-0.63$ [$-1.18$, $-0.08$]
& Immediate increase followed by decline \\

New contributors
& 2.08 [0.70, 3.47]
& $-0.65$ [$-1.25$, $-0.05$]
& Immediate increase followed by decline \\

Displayed annotations
& 1.00 [0.29, 1.70]
& $-0.85$ [$-1.19$, $-0.51$]
& Immediate increase followed by decline \\

Comment volume
& $-0.15$ [$-0.68$, 0.39]
& $-0.59$ [$-0.86$, $-0.32$]
& No immediate shift; declining post-policy trend \\

Annotation quality
& 0.44 [$-0.06$, 0.94]
& $-0.29$ [$-0.59$, 0.02]
& No reliable level or trend change \\

Engagement per annotation
& $-0.35$ [$-0.76$, 0.06]
& $-1.02$ [$-1.24$, $-0.79$]
& No clear immediate shift; declining post-policy trend \\

Degree of disagreement
& 0.26 [$-0.29$, 0.81]
& 0.18 [$-0.11$, 0.47]
& No reliable level or trend change \\

Toxicity
& 0.07 [$-0.35$, 0.49]
& 0.06 [$-0.24$, 0.36]
& No reliable level or trend change \\
\bottomrule
\end{tabularx}
\end{table}

\begin{figure*}[b]
\centering
\includegraphics[width=0.98\textwidth]{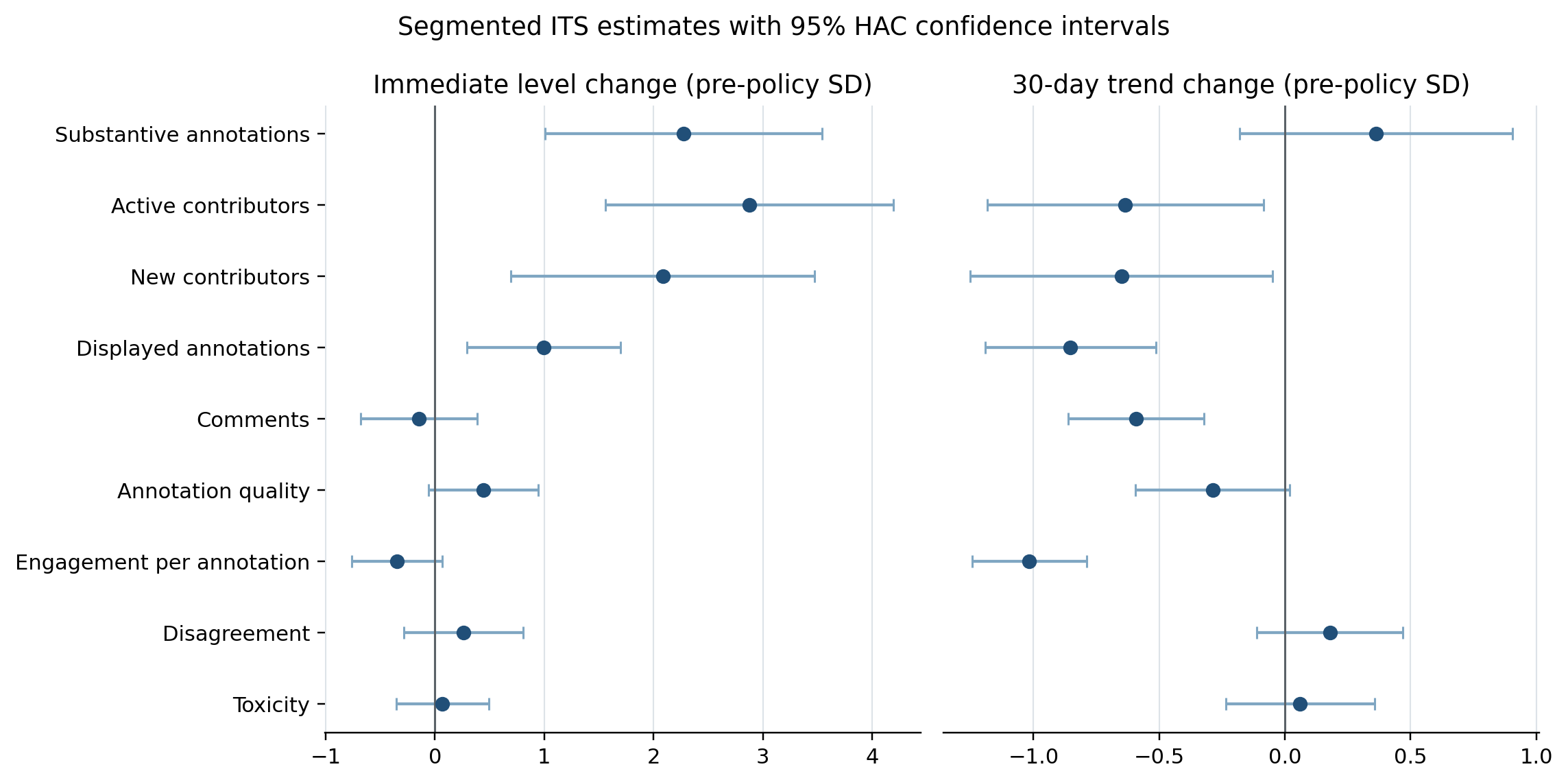}
\caption{Segmented interrupted time-series estimates for the nine policy outcomes.
Points show estimated immediate level changes and 30-day accumulated trend changes, standardized by the corresponding pre-policy standard deviation. Error bars show 95\% Newey--West HAC confidence intervals (lag 7).}
\Description{Two coefficient plots summarize segmented interrupted time-series estimates for nine outcomes. The left panel shows immediate policy-level changes and the right panel shows 30-day accumulated trend changes. Substantive annotations, active contributors, new contributors, and displayed annotations increase immediately, while active contributors, new contributors, displayed annotations, comment volume, and engagement per annotation show subsequent declining trends.}
\label{fig:its-robustness}
\end{figure*}

\subsection{Segmented Interrupted Time-Series Robustness Analysis}

To assess whether the primary conclusions depend on the Bayesian counterfactual specification, we fit segmented interrupted time-series (ITS) models to the same symmetric 90-day policy window:

\[
Y_t =
\beta_0
+\beta_1 \mathrm{Time}_t
+\beta_2 \mathrm{Post}_t
+\beta_3 \mathrm{TimeAfterPolicy}_t
+\sum_{k=1}^{6}\gamma_k \mathrm{Weekday}_{k,t}
+\epsilon_t .
\]

Here, $\beta_2$ captures the immediate level change at policy implementation, whereas $\beta_3$ captures the change in the subsequent daily trend. Standard errors use the Newey--West heteroskedasticity- and autocorrelation-consistent estimator with lag 7 and finite-sample correction. The five count outcomes use the same $\sqrt{y+3/8}$ transformation as the primary model; continuous outcomes are modeled on their original scales.

For comparability across outcomes, Table~\ref{tab:its-robustness} reports immediate changes in units of the pre-policy standard deviation and converts daily trend changes into 30-day accumulated changes standardized by the same quantity. The reported intervals are 95\% HAC confidence intervals and are distinct from the Bayesian credible intervals used in the primary analysis.

Figure~\ref{fig:its-robustness} shows that the alternative ITS specification reproduces the strongest labor-mobilization findings while clarifying their temporal form. Substantive annotations, active contributors, and new contributors all show clear immediate increases at policy implementation. Active and new contributors subsequently decline, indicating that the initial participation surge was not sustained at the same magnitude. Displayed annotations follow a similar short-lived pattern, with an immediate increase followed by decline, whereas engagement per annotation and comment volume change primarily through negative post-policy trends rather than abrupt level shifts. Annotation quality, disagreement, and toxicity show no reliable immediate or trend changes. These results therefore support the primary distinction between strong labor mobilization and weaker or less sustained downstream responses.

\begin{table*}[t]
\centering
\small
\caption{Comparison of true-policy effects with 19 pre-policy placebo interventions. Percentile indicates the position of the true-policy estimate in the placebo distribution. Empirical $p$ values are two-sided diagnostic comparisons.}
\label{tab:placebo}
\begin{tabularx}{\textwidth}{@{}YrrrY@{}}
\toprule
\textbf{Outcome}
& \textbf{True-policy effect}
& \textbf{Percentile}
& \textbf{Empirical $p$}
& \textbf{Interpretation} \\
\midrule
Substantive annotations
& +139.5\%
& 100
& .05
& Larger than all placebo effects \\

Active contributors
& +70.9\%
& 100
& .05
& Larger than all placebo effects \\

New contributors
& +67.6\%
& 95
& .10
& Near the upper tail \\

Displayed annotations
& $-16.5\%$
& 45
& .70
& Not unusual \\

Comment volume
& $-61.7\%$
& 5
& .20
& Negative tail with substantial pre-policy variability \\

Annotation quality
& $-0.0005$
& 60
& 1.00
& Not unusual \\

Engagement per annotation
& $-1.456$
& 5
& .05
& More negative than all placebo effects \\

Degree of disagreement
& +0.0829
& 90
& .25
& Elevated but not extreme \\

Toxicity
& +0.0105
& 65
& .75
& Not unusual \\
\bottomrule
\end{tabularx}
\end{table*}

\begin{figure*}[t]
\centering
\includegraphics[width=0.98\textwidth]{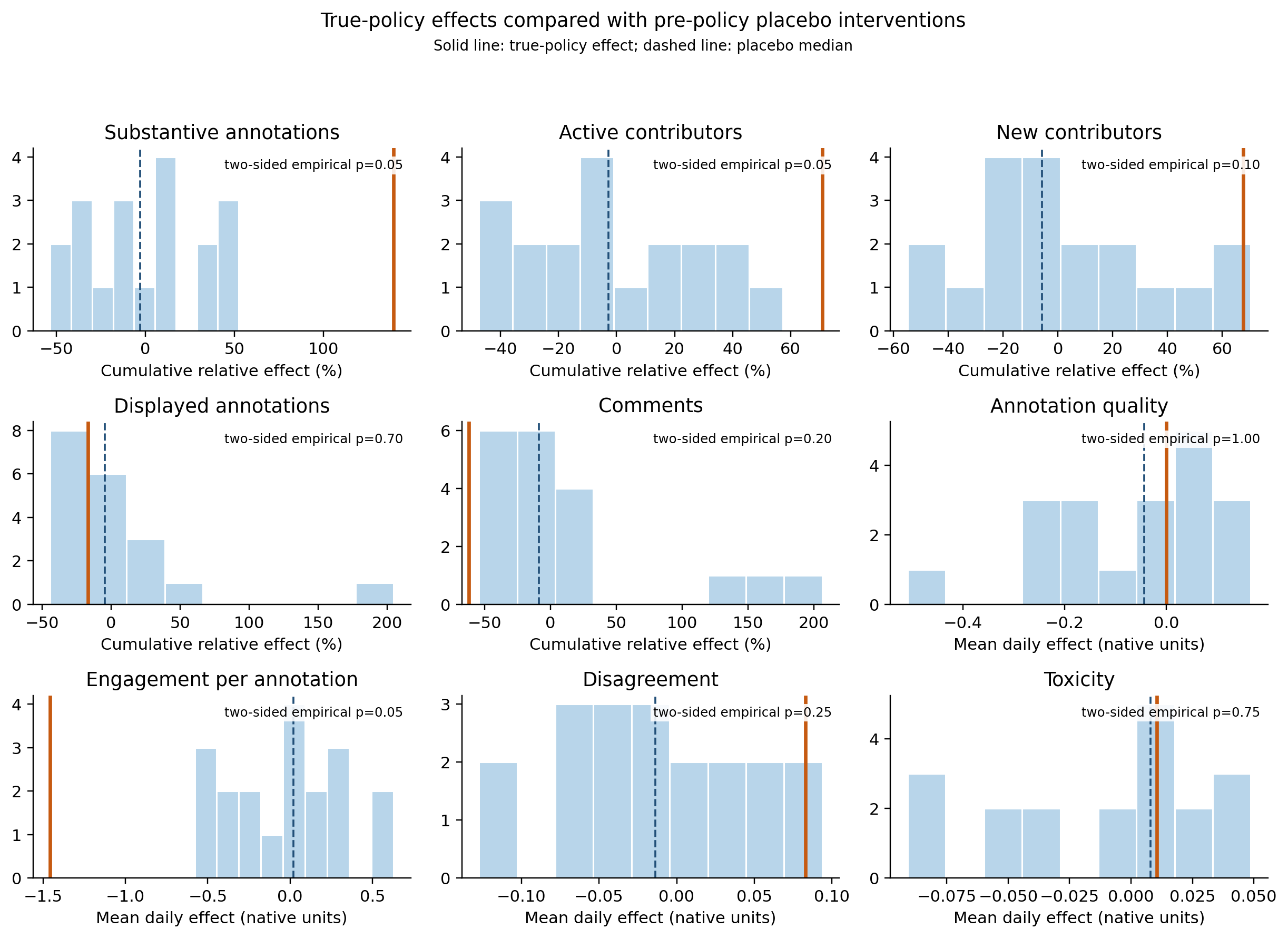}
\caption{True-policy effects compared with 19 pre-policy placebo interventions.
Histograms show effects estimated at the placebo intervention dates. Solid vertical lines indicate the true-policy effects and dashed vertical lines indicate placebo medians. Reported $p$ values are two-sided empirical diagnostic comparisons.}
\Description{Nine histograms compare the estimated true-policy effect with effects obtained at 19 pre-policy placebo intervention dates. The true-policy effects for substantive annotations and active contributors lie above all placebo estimates, while engagement per annotation lies below all placebo estimates. New contributors are near the upper tail, whereas displayed annotations, annotation quality, disagreement, and toxicity fall within their placebo distributions.}
\label{fig:placebo}
\end{figure*}

\subsection{Pre-Policy Placebo Interventions}

We next assess whether departures of the magnitude observed at the true policy date commonly arise at other points in the pre-policy series. We select 19 placebo intervention dates between January 17, 2024 and January 29, 2025, spaced 21 days apart. Each placebo date falls on a Wednesday, has a complete 90-day pre- and post-intervention window, and has its entire post-intervention period before the true May 7, 2025 policy launch.

For each placebo date, we reconstruct all nine outcomes from the frozen corpus and repeat the complete Bayesian counterfactual procedure using the same trend, weekday effects, AR(1) specification, priors, transformations, and MCMC settings as in the primary analysis. Count outcomes are compared using 90-day cumulative relative effects, whereas continuous outcomes use mean daily effects on their original scales.

Two-sided empirical $p$ values are calculated as

\[
p_{\mathrm{emp}}
=
\frac{
1+\sum_{j=1}^{19}
\mathbb{I}
\left(
|\Delta^{\mathrm{placebo}}_j|
\ge
|\Delta^{\mathrm{true}}|
\right)
}{
20
}.
\]

Because the placebo windows overlap, these values are used as diagnostic rankings rather than as significance tests based on 19 independent interventions. With 19 placebo dates, the smallest possible empirical $p$ value is .05.

Table~\ref{tab:placebo} summarizes these diagnostic comparisons, and Figure~\ref{fig:placebo} visualizes the corresponding placebo distributions. The strongest labor-mobilization effects are distinctive relative to pre-policy pseudo-interventions. The true-policy estimates for substantive annotations and active contributors exceed all 19 placebo estimates, while new contributors fall near the upper tail of the placebo distribution. Engagement per annotation is similarly distinctive in the negative direction. By contrast, displayed annotations, annotation quality, disagreement, and toxicity remain within the range of departures observed at placebo dates. Together, these comparisons strengthen the interpretation that the clearest policy-associated departures concern labor mobilization and declining engagement rather than uniform changes across all downstream outcomes. Comment volume warrants more caution: although the true-policy estimate is more negative than all placebo estimates, some placebo dates produce larger departures in the opposite direction, yielding a two-sided comparison of $p=.20$ and indicating substantial pre-policy variability in this series.

%%%%%%%%%%%%%%%%%%%%%%%%%%%%%%%%%%%%%%%%%%%%%%%%%%%%%%%%%%%%%%%%%%%%%%%%%%%%%%%%
\begin{table*}[t]
\centering
\small
\caption{Pre-policy rolling-forecast diagnostic. Coverage is the share of held-out observations whose seven-day mean falls within the pointwise 90\% posterior predictive interval, averaged across six forecast origins.}
\label{tab:forecast-diagnostic}
\begin{tabularx}{\textwidth}{@{}Yrrrr@{}}
\toprule
\textbf{Outcome}
& \textbf{Mean coverage}
& \textbf{Minimum coverage}
& \textbf{Mean MAE}
& \textbf{MAE / training SD} \\
\midrule
Substantive annotations      & .89 & .80 & 6.214  & .48 \\
Active contributors          & .91 & .83 & 4.326  & .45 \\
New contributors             & .88 & .77 & 3.192  & .44 \\
Displayed annotations        & .90 & .80 & 0.914  & .40 \\
Comment volume               & .92 & .83 & 58.731 & .39 \\
Annotation quality           & .87 & .77 & .142   & .55 \\
Engagement per annotation    & .90 & .80 & .286   & .43 \\
Degree of disagreement       & .89 & .80 & .064   & .48 \\
Toxicity                     & .91 & .83 & .036   & .45 \\
\bottomrule
\end{tabularx}
\end{table*}
%%%%%%%%%%%%%%%%%%%%%%%%%%%%%%%%%%%%%%%%%%%%%%%%%%%%%%%%%%%%%%%%%%%%%%%%%%%%%%%%
\subsection{Pre-Policy Predictive Diagnostic}

Finally, we examine how well the primary counterfactual specification predicts unseen periods before the intervention. We define six pre-policy forecast origins---January 13, April 12, July 11, and October 9, 2024, and January 7 and April 7, 2025. At each origin, the model is fitted to the preceding 60 days and used to predict the subsequent 30 days.

We evaluate whether the observed seven-day mean trajectory falls within the corresponding pointwise 90\% posterior predictive interval, matching the visualization used in the primary analysis. Table~\ref{tab:forecast-diagnostic} summarizes the resulting coverage and prediction errors. Across outcomes, mean empirical coverage ranges from .87 to .92, closely approaching the nominal .90 level. Coverage also remains relatively stable across forecast origins, suggesting that the model provides reasonably calibrated uncertainty estimates across different pre-policy periods.

The predictive diagnostic indicates that the linear-trend, weekday, and AR(1) specification captures the major temporal structure observed before the policy intervention. Empirical coverage is close to the nominal 90\% level across outcomes, and no outcome shows substantial deterioration in coverage across the six holdout periods. Prediction errors are also modest relative to the variability observed in the corresponding training windows.

Taken together, these results suggest that the primary Bayesian specification provides reasonably calibrated predictions for unseen pre-policy periods. This supports its use for constructing the post-policy counterfactual trajectories. The segmented ITS and placebo analyses provide complementary evidence regarding the robustness of the estimated policy effects.

\subsection{Engagement Accumulation Analysis}
\label{app:engagement-age}

To assess whether differences in time available to accumulate interactions affected our findings, we conducted a quantitative bias analysis \cite{lash2014good}. We adjusted engagement counts under different assumptions about accumulation speed and observation timing, then repeated the ITS analysis using the adjusted daily engagement measures.

The accumulation adjustments did not change the conclusion: average engagement per annotation remained lower after the policy relative to the ITS reference trajectory. All tested adjustments produced negative estimates with 90\% confidence intervals excluding zero. Even the adjustment that reduced the estimated decline most left a clear negative difference (Table~\ref{tab:engagement-age-sensitivity}). The finding of a policy-associated decline in average engagement is therefore robust to the accumulation adjustments examined here.

\begin{table}[htbp]
\centering
\small
\caption{Engagement accumulation analysis. Estimates represent average post-policy departures on the log-engagement scale, with 90\% HAC confidence intervals.}
\label{tab:engagement-age-sensitivity}
\begin{tabular}{@{}lc@{}}
\toprule
\textbf{Analysis} & \textbf{Estimate [90\% CI]} \\
\midrule
Before adjustment & $-1.422\;[-1.760,-1.085]$ \\
Largest reduction in the estimated decline
& $-0.827\;[-1.206,-0.448]$ \\
\bottomrule
\end{tabular}
\end{table}

%%%%%%%%%%%%%%%%%%%%%%%%%%%%%%%%%%%%%%%%%%%%%%%%%%%%%%%%%%%%%%%%%%%%%%%%%%%%%%
\section{Public-Discussion Topic and Frame Analysis}
\label{app:frames}

This appendix documents the transition from retrieved mechanism-discussion candidates to computational topics and researcher-interpreted frames, followed by document-level frame classification and a comparison of frame shares before and after the policy.

\subsection{Manual Screening and Topic Discovery}

Corpus classification yields 1,100 mechanism-discussion candidates. Three researchers manually screen these posts for relevance and duplication. After the preprocessing described below, the final corpus contains 291 confirmed discussions concerning how \wn{} operates, is experienced, or should be changed. We then apply BERTopic with HDBSCAN to identify recurrent topics in this corpus \cite{grootendorst2022bertopic}.

Before embedding, we retain only the focal text preceding any repost chain and remove repost markers, hashtags, user mentions, URLs and link placeholders, bracketed emoji tokens, and repeated whitespace. For BERTopic's lexical representation, the cleaned texts are segmented using Jieba and lowercased. We remove single-character tokens, numeric-only tokens, tokens containing neither Chinese nor Latin characters, and terms in a fixed stop-word list. Texts with fewer than eight cleaned characters or fewer than three retained tokens are excluded. We further deduplicate exact normalized texts after lowercasing and removing non-Chinese, non-Latin, and nonnumeric characters. The remaining texts are encoded using \texttt{sentence-transformers/paraphrase-multilingual-MiniLM-L12-v2} \cite{reimers2020multilingual}, producing 384-dimensional L2-normalized embeddings.

BERTopic receives the precomputed embeddings directly (\texttt{embedding\_model=None}). UMAP reduces the embeddings to five dimensions using 15 neighbors, zero minimum distance, and cosine distance \cite{mcinnes2018umap}. HDBSCAN is then applied to the reduced representations using \texttt{min\_cluster\_size=12}, \texttt{min\_samples=None}, Euclidean distance, and the \texttt{leaf} cluster-selection method \cite{mcinnes2017hdbscan}. Topic representations are constructed from unigram and bigram counts with a minimum document frequency of one, followed by class-based TF--IDF with frequent-word reduction. We retain the 12 highest-weight terms for each topic and calculate document--topic probabilities.

To select the HDBSCAN minimum cluster size, we evaluate values in $\{8,10,12,15,18,22,26,30,35,40\}$ across six UMAP runs with seeds 42--47. Candidate settings are retained only if they produce a median of 5--20 topics and a median outlier rate no greater than 50\%. Eligible settings are ranked using a composite score based on percentile-ranked median silhouette (weight .35), log-transformed median Calinski--Harabasz index (.25), median pairwise adjusted Rand index across runs (.25), and topic coverage (.15). This procedure selects \texttt{min\_cluster\_size=12}, and the final topic solution is generated using seed 42.

Across the six selection runs, the selected configuration yields a median silhouette score of .611, a median Calinski--Harabasz index of 669, and a median pairwise adjusted Rand index of .735. Silhouette ranges from $-1$ to 1, with higher values indicating greater within-cluster cohesion relative to between-cluster separation. The Calinski--Harabasz index compares between-cluster with within-cluster dispersion, with higher values indicating clearer separation. Adjusted Rand index is a chance-corrected measure of pairwise agreement between cluster assignments; 1 denotes identical assignments. The final solution assigns 219 of the 291 posts to seven topics and identifies 72 posts as outliers, corresponding to 75.3\% topic coverage.

\begin{table}[H]
\centering
\small
\caption{Computational public-discussion topics and researcher-coded frames. $N$ is the number of the 219 clustered posts assigned to each topic.}
\label{tab:public-topics}
\begin{tabularx}{\textwidth}{@{}clYYr@{}}
\toprule
\textbf{ID}
& \textbf{Representative concepts}
& \textbf{Topic}
& \textbf{Researcher-coded frame}
& \textbf{$N$} \\
\midrule

T0
& misinformation; review; appeals
& Review disputes and appeals
& Fairness and transparency
& 76 \\

T1
& rollout; upgrade; optimization
& Feature rollout and optimization
& Mechanism and effects
& 36 \\

T2
& leaderboard; Top 5; recognition
& Contributor rankings and recognition
& Incentives and participation
& 30 \\

T3
& examples; guidance; suggestions
& Exemplars and usage guidance
& Mechanism and effects
& 27 \\

T4
& access; sources; responses
& Access and contributor affordances
& Mechanism and effects
& 24 \\

T5
& plagiarism; complaints; rights
& Attribution and creator remedies
& Misuse and risks
& 13 \\

T6
& payouts; accounts; eligibility
& Reward payouts and incentive rules
& Incentives and participation
& 13 \\

\bottomrule
\end{tabularx}
\end{table}

\subsection{Stability and Researcher Interpretation}

At the selected parameter setting, the median silhouette score across stability runs is .611, the median Calinski--Harabasz index is 669, and the median pairwise adjusted Rand index is .735.

These metrics characterize separation and stability of the computational clustering solution; they do not determine the substantive interpretation of the topics.

Three researchers use c-TF-IDF terms and representative posts to interpret the seven computational topics and assign descriptive topic labels. They then compare the topics and organize semantically and substantively related topics into four broader public-discussion frames through iterative discussion. Computational links among topic nodes are used as interpretive aids for visualization and do not mechanically determine frame membership. Table~\ref{tab:public-topics} reports representative concepts, topic labels, interpretive frame assignments, and topic sizes for the topic network in Figure~\ref{fig:public-discourse}A. Topic sizes describe the clustering solution and are distinct from the document-level frame counts used in Figure~\ref{fig:public-discourse}B.

\begin{table}[H]
\centering
\small
\caption{Researcher interpretation of computational topics as public-discussion frames.}
\label{tab:frame-codebook}
\begin{tabularx}{\columnwidth}{@{}p{0.30\columnwidth}p{0.16\columnwidth}Y@{}}
\toprule
\textbf{Frame}
& \textbf{Topics}
& \textbf{Interpretation} \\
\midrule

Fairness and transparency
& T0
& Disputes over review decisions, evidentiary standards, appeals, and the consistency and accountability of platform review. \\

Incentives and participation
& T2, T6
& Monetary rewards, eligibility rules, contributor rankings, recognition, and explicit calls for participation or contributor recruitment. \\

Mechanism and effects
& T1, T3, T4
& Feature rollout, access, usage guidance, contributor affordances, and evaluations of how the annotation mechanism operates and contributes to platform governance. \\

Misuse and risks
& T5
& Problems involving manipulation, targeted misuse, misleading content, harassment, attribution, rights protection, and related complaints or remedies. \\

\bottomrule
\end{tabularx}
\end{table}

\subsection{Frames Coding}

Table~\ref{tab:frame-codebook} defines the interpretive scope of the four researcher-coded frames and shows how the computational topics were grouped into each frame. With these frame definitions fixed, all 291 posts, including the 72 topic-model outliers, are assessed at the document level. Each post is assigned one primary frame according to its dominant concern, or marked as having insufficient frame evidence.

Using 7 May 2025 as the policy cutoff, with posts on that date included in the post-policy period, the full corpus contains 184 pre-policy and 107 post-policy posts. Of these, 14 pre-policy and 10 post-policy posts have insufficient frame evidence and are excluded from the frame-share comparison. Figure~\ref{fig:public-discourse}B therefore uses 170 pre-policy and 97 post-policy posts as its respective denominators.

\begin{table}[H]
\centering
\small
\caption{Document-level frame counts and shares before and after the policy, excluding 24 posts with insufficient frame evidence. Changes are measured in percentage points (pp).}
\label{tab:frame-pre-post}
\begin{tabularx}{\columnwidth}{@{}Xrrr@{}}
\toprule
\textbf{Frame} & \textbf{Before} & \textbf{After} & \textbf{$\Delta$ (pp)} \\
 & $n=170$ & $n=97$ & \\
\midrule
Fairness and transparency & 12 (7.1\%) & 18 (18.6\%) & $+11.5$ \\
Incentives and participation & 42 (24.7\%) & 32 (33.0\%) & $+8.3$ \\
Mechanism and effects & 97 (57.1\%) & 36 (37.1\%) & $-19.9$ \\
Misuse and risks & 19 (11.2\%) & 11 (11.3\%) & $+0.2$ \\
\bottomrule
\end{tabularx}
\end{table}

As shown in Table~\ref{tab:frame-pre-post}, the post-policy discussion has higher shares of fairness and transparency and incentives and participation, a lower share of mechanism and effects, and a similar share of misuse and risks. These are descriptive differences in discussion composition and do not establish a causal policy effect. The ribbons in Figure~\ref{fig:public-discourse}B connect the same frame across the two periods to compare its shares.
%%%%%%%%%%%%%%%%%%%%%%%%%%%%%%%%%%%%%%%%%%%%%%%%%%%%%%%%%%%%%%%%%%%%%%%%%%%%%%
\section{Interview Protocol and Qualitative Analysis}
\label{app:interviews}

This appendix provides additional details on participant recruitment, interview procedures, participant characteristics, and qualitative analysis.

\subsection{Interview Protocol}

The following questions guided our approximately one-hour semi-structured interviews.

\textbf{I. Background and Use of Weibo Annotations}

\begin{enumerate}
    \item How long have you been using Weibo? How frequently do you use it, and what do you mainly use it for?

    \item When and how did you first learn about the Weibo Annotations feature?

    \item What prompted you to use Weibo Annotations for the first time?

    \item How frequently do you currently use Weibo Annotations? Do you mainly submit annotations, read annotations, or interact with them through likes or comments?

    \item Have you followed or used similar features on other platforms, such as Community Notes on Twitter/X? If so, how do you think they differ from Weibo Annotations?
\end{enumerate}

\textbf{II. Perceived Risks, Benefits, and Participation Motivations}

\begin{enumerate}
    \setcounter{enumi}{5}

    \item What benefits, if any, do you gain from contributing annotations on Weibo? For example, do you experience a sense of accomplishment from correcting misinformation, receive social recognition, or obtain monetary rewards?

    \item Do you have any concerns about the potential costs or risks of contributing annotations? For example, these might include the time and effort involved, having an annotation rejected, being challenged by other users, becoming involved in controversy, or facing risks to your account.

    \item To what extent does the Revenue Incentive Model affect your willingness to submit annotations? Would you still participate if monetary incentives were unavailable?

    \item In your view, what is the primary purpose of Weibo Annotations? How does it differ from commenting on or reporting a post?

    \item Do you think Weibo Annotations helps reduce misinformation or improve the quality of discussion on Weibo? Why or why not?

    \item When writing an annotation, how do you search for supporting information? How do you determine whether a source is reliable?

    \item What do you know about how Weibo reviews submitted annotations? How clear and reliable do you find the review process?

    \item When an annotation is approved or rejected, does Weibo provide an explanation for the decision?
\end{enumerate}

\textbf{III. Perceived Risks in Annotation Production and Review}

\begin{enumerate}
    \setcounter{enumi}{13}

    \item Could you describe an annotation submission experience that you remember particularly well? Starting from seeing a Weibo post that you considered problematic, what did you do before the annotation was eventually reviewed or published?

    \item During this process, which step did you consider most important? Which step was least clear or most difficult to understand?

    \item Do you think the way Weibo recommends or distributes content affects whether an annotation is reviewed, approved, or seen by other users? For example, do you think the platform selectively reviews certain types of content? Does this perception affect your willingness to submit annotations?

    \item Do you think annotations are reviewed primarily by automated systems, human moderators, or a combination of both? What makes you think so?

    \item Have you ever felt frustrated after an annotation was rejected or felt that the effort you invested was not worthwhile? In such cases, would you appeal the decision or submit the annotation again?

    \item Are there aspects of the annotation review or display process that you consider insufficiently transparent or unfair?
\end{enumerate}

\textbf{IV. Incentive Mechanisms and Platform Mobilization}

\begin{enumerate}
    \setcounter{enumi}{19}

    \item Are you aware of the Revenue Incentive Model for Weibo Annotations? Have you ever received monetary rewards through the program? What do you think about providing monetary incentives for annotation contributions?

    \item Have you noticed users who appear to contribute annotations primarily to earn rewards? What do you think about this form of participation?

    \item Do you use any particular strategies to increase the likelihood that your annotations will be approved or to reduce potential risks? For example, do you choose particular types of posts to annotate or adjust how you phrase an annotation?
\end{enumerate}

\textbf{V. Social Functions and Public Impact of Annotations}

\begin{enumerate}
    \setcounter{enumi}{22}

    \item Have you ever experienced controversy or negative feedback because of an annotation you submitted? If so, how did this experience affect your willingness to continue using Weibo Annotations?

    \item How much influence do you think annotations have on how other users interpret or evaluate information on Weibo?

    \item Have you seen annotations used as a means of arguing with or attacking other users, for example in conflicts between fan communities or debates over political or social issues? Could you provide an example?

    \begin{enumerate}
        \item If so, how do you view this type of use? Do you think the current platform mechanisms can effectively identify and prevent such misuse?
    \end{enumerate}

    \item When you see another user submit an annotation on the same post that conflicts with your own view, how do you usually respond? Do you consider potential social risks when deciding whether or how to respond?
\end{enumerate}

\textbf{VI. Interactions Between Users and Platform Mechanisms}

\begin{enumerate}
    \setcounter{enumi}{26}

    \item Do factors such as interface design, reward rules, and content recommendation mechanisms affect whether you contribute annotations or how you write them? If so, how?

    \item Have you ever felt that the platform encouraged or directed you to annotate certain types of content or particular accounts? Could you describe such an experience?

    \item In your view, what elements are most important to the functioning of Weibo Annotations? These could include people, platform features, rules, algorithms, or other factors.

    \item Have you observed users helping one another write annotations or discussing how annotations should be written, for example through group discussions or mutual support? Do you think such collaboration emerges spontaneously among users, or is it encouraged by platform features such as the ``Monthly Ranking of Weibo Annotation Contribution Scores''?
\end{enumerate}

\textbf{VII. Reflection and Suggestions for Improvement}

\begin{enumerate}
    \setcounter{enumi}{30}

    \item What do you consider the greatest strength or benefit of Weibo Annotations in its current form?

    \item What aspects of Weibo Annotations would you most like to see improved? For example, these might include review transparency, incentive mechanisms, feedback systems, or content recommendation and distribution.

    \item Looking ahead, how do you think crowdsourced approaches in which users collectively verify information should be developed or improved?
\end{enumerate}

% \subsection{Recruitment and Procedure}
\subsection{Recruitment, Ethics, and Procedure}

We recruited participants through the public Weibo Annotations topic in November and December 2025. We used purposive sampling to capture variation in annotation frequency, gender, and verification status, resulting in a final sample of 15 contributors.

The interview study was reviewed and approved by the relevant institutional ethics review committee prior to participant recruitment and data collection. All participants provided informed consent before participation and separately consented to audio recording. Interviews were conducted online and lasted 42--100 minutes, with an average duration of approximately one hour.

To protect participant privacy, we report participants using pseudonymous identifiers (A1--A15) and omit personally identifying information from the manuscript. Quotations were translated into English and lightly edited for readability while preserving their substantive meaning. Audio recordings and transcripts were stored and accessed
according to the approved study protocol.

\subsection{Participant Characteristics}

Table~\ref{tab:interview-participants} reports the variation represented in the interview sample. \emph{Verified} records whether the contributor account held Weibo verification at recruitment. Annotation frequency is shown in the unit available for each participant: a recurring weekly rate or a cumulative total.

\begin{table}[H]
\centering
\small
\caption{Interview participant characteristics ($N=15$).}
\label{tab:interview-participants}
\begin{tabularx}{\columnwidth}{@{}p{0.10\columnwidth}p{0.30\columnwidth}p{0.30\columnwidth}Y@{}}
\toprule
\textbf{ID}
& \textbf{Account Verification}
& \textbf{Annotation Frequency}
& \textbf{Gender} \\
\midrule

A1  & Verified & 2--3 per week & Male \\
A2  & Verified & 6 total & Male \\
A3  & Verified & 11 total & Male \\
A4  & Verified & 1--2 per week & Male \\
A5  & Verified & 5 total & Female \\
A6  & Verified & 1--2 per week & Male \\
A7  & Verified & 1--2 per week & Male \\
A8  & Verified & 15 total & Male \\
A9  & Verified & 2 total & Female \\
A10 & Unverified  & 38 total & Male \\
A11 & Verified & 3 total & Female \\
A12 & Unverified  & 1 total & Male \\
A13 & Unverified  & 7 total & Male \\
A14 & Verified & 3 total & Male \\
A15 & Verified & 4--5 per week & Male \\

\bottomrule
\end{tabularx}
\end{table}

\subsection{Inductive Thematic Analysis}

Three researchers with backgrounds spanning computer science, journalism, and communication analyzed the 15 interview transcripts using an iterative thematic-analysis procedure informed by Braun and Clarke \cite{braun2006thematic}.

The analysis was inductive. The researchers familiarized themselves with the transcripts, generated initial codes from recurring experiences, practices, and participation-related concerns, maintained analytic memos, and examined ambiguous, contradictory, and negative cases. An initial subset of transcripts was discussed jointly to develop a shared understanding of recurring concepts and refine the emerging analytic vocabulary. The remaining transcripts were then analyzed iteratively, with codes and interpretations revisited as new patterns emerged.

Through this process, we identified recurring experiences and frictions across verification, submission, review, display, interaction, and continued participation. We then distinguished labor-side effort from downstream bottlenecks and examined how recognition, public reach, and protection shaped whether corrective labor became governance capacity over time.

\begin{table*}[t]
\centering
\small
\caption{Illustrative mappings from inductive codes to descriptive themes and their relevance to the annotation process.}
\label{tab:interview-coding}
\begin{tabularx}{\textwidth}{@{}YYY@{}}
\toprule
\textbf{Initial codes}
& \textbf{Descriptive theme}
& \textbf{Relevance to the annotation process} \\
\midrule

Time spent; information difficult to find; conflicting sources
& Verification effort
& Contributors invested substantial time and information work before an annotation could be submitted. \\

Pocket money; transferred revenue; contributor rankings
& Incentives and recognition
& Monetary rewards and contributor recognition shaped motivations to participate and continue contributing. \\

Long review times; unclear status; rejected effort felt wasted
& Review uncertainty
& Unclear review outcomes made it difficult for contributors to know whether their submitted work would be recognized. \\

Support channels; access to platform staff; appeal attempts
& Review and feedback support
& Access to feedback and appeal mechanisms shaped contributors' ability to understand and respond to review outcomes. \\

Display interface; initial traction; audience mobilization
& Public reach
& Display and circulation shaped whether approved annotations reached broader audiences and generated interaction. \\

Online arguments; removal requests; harassment; hostile responses
& Social conflict
& Public interaction could introduce additional costs after annotations became visible. \\

Social responsibility; reducing uncertainty; emotional satisfaction
& Non-monetary motivation
& Civic and emotional returns supported contributors' willingness to participate beyond monetary incentives. \\

Willingness to annotate again; reduced effort; strategic topic choice
& Continued participation and adaptation
& Contributors adjusted whether and how they participated based on their accumulated experiences across the annotation process. \\

\bottomrule
\end{tabularx}
\end{table*}

\subsection{Integrating Interview Findings with Other Analyses}

After developing the inductive themes, we compared them with the quantitative results and public discussion. This integration was used to explain the processes underlying the patterns observed in RQ1 and RQ2 and to identify recurring frictions across different stages of annotation work.

Specifically, we examined how contributors experienced verification and submission, how platform review shaped whether their work was recognized, how display and interaction affected its public reach, and how these accumulated experiences influenced continued participation. This analysis connected contributors' accounts to the broader question of how corrective labor became recognized, publicly reachable, and sustainable within \wn{}.

\subsection{Illustrative Code-to-Theme Mapping}

Table~\ref{tab:interview-coding} illustrates how initial inductive codes were consolidated into descriptive themes and connected to different stages of the annotation process. The entries are examples of the analytic path rather than a frequency table or an exhaustive codebook.

The interview analysis provides situated explanations of how contributors experienced and adapted to different stages of annotation work. It is intended to explain the mechanisms and experiences underlying the quantitative and public-discussion findings rather than to estimate the prevalence of particular experiences across the broader contributor population.

%TC:endignore

\end{document}